\documentclass[a4paper,11pt,reqno]{amsart}
\usepackage{eurosym}
\usepackage[a4paper,hscale=0.7,vscale=0.75,centering]{geometry}
\usepackage{stmaryrd}
\usepackage{bm}
\usepackage{amsthm}
\usepackage{hyperref}
\usepackage{amssymb}
\usepackage{latexsym}
\usepackage{amsmath}
\usepackage{amsfonts}
\usepackage{mathabx}
\usepackage[dvips]{graphicx}
\usepackage[utf8]{inputenc}
\usepackage[LGR,T1]{fontenc}
\usepackage[francais,english]{babel}
\usepackage{url}
\usepackage{enumerate}
\usepackage{hyperref}
\usepackage[usenames,dvipsnames]{color}
\usepackage{mathtools}
\usepackage{fancyhdr}
\usepackage[toc,page]{appendix}
\usepackage{chngcntr}
\newtheorem{Theorem}{Theorem}
\newtheorem{theorem}{Theorem}
\newtheorem{Definition}[Theorem]{Definition}
\newtheorem{Proposition}[Theorem]{Proposition}

\newtheorem{Assumption}{Assumption}
\newtheorem{Lemma}[Theorem]{Lemma}
\newtheorem{Corollary}[Theorem]{Corollary}
\newtheorem{Remark}[Theorem]{Remark}

\makeatletter \@addtoreset{equation}{section}
\def\theequation{\thesection.\arabic{equation}}
\@addtoreset{Theorem}{section}

\@addtoreset{Assumption}{section}
\newcommand{\dif}{\mathop{}\!\mathrm{d}}
\input{tcilatex}

\newcommand{\cE}{\mathcal{E}}
\newcommand{\cF}{\mathcal{F}}

\newcommand{\cH}{\mathcal{H}}
\newcommand{\cI}{\mathcal{I}}
\newcommand{\cJ}{\mathcal{J}}
\newcommand{\cK}{\mathcal{K}}
\newcommand{\cL}{\mathcal{L}}
\newcommand{\cM}{\mathcal{M}}

\newcommand{\cS}{\mathcal{S}}

\renewcommand{\P}{\mathbb{P}}

\newcommand{\R}{\mathbb{R}}

\def \proof{{\noindent \bf Proof. }}
\def \eproof{\hbox{ }\hfill$\Box$}

\newcommand{\ud}{\mathrm{d}}

\newcommand{\1}{{\bf 1}}
    
\newcommand{\set}[1]
    {\ensuremath{\{ #1 \}}}
    
\newcommand{\HP}[1] 
    {\ensuremath{\mathscr{H}^{#1}}}

\newcommand{\esp}[1]{\ensuremath{\mathbb{E} \!\! \left[#1\right] }}

\renewcommand{\Xi}[1]{X_{i #1}}

\begin{document}
\title[Singular FBSDEs and multi-period carbon markets]{Modelling
multi-period carbon markets using singular forward backward SDEs}
\author[Jean-Fran\c{c}ois Chassagneux, Hinesh Chotai and Dan Crisan]{%
Jean-Fran\c{c}ois Chassagneux$^*$, Hinesh Chotai$^{\dagger}$ and Dan Crisan$%
^{\ddagger}$ }

\begin{abstract}
We introduce a model for the evolution of emissions and the price of emissions
allowances in a carbon market such as the European Union Emissions Trading
System (EU ETS). The model accounts for multiple trading periods, or phases, with multiple times at which compliance can occur. At the end of
each trading period, the participating firms must surrender allowances for
their emissions made during that period, and additional allowances can be
used for compliance in the following periods. We show that the multi-period
allowance pricing problem is well-posed for various mechanisms (such as
banking, borrowing and withdrawal of allowances) linking the trading periods. The results
are based on the analysis of a forward-backward stochastic differential
equation with coupled forward and backward components, a discontinuous terminal condition and a forward component that is degenerate. We also introduce an infinite period model, for
a carbon market with a sequence of compliance times and with no end date. We show
that, under appropriate conditions, the value function for the multi-period
pricing problem converges, as the number of periods increases, to a value
function for this infinite period model, and that such functions are unique.
\end{abstract}

\maketitle

\renewcommand{\thefootnote}{\fnsymbol{footnote}} \footnotetext[1]{%
Laboratoire de Probabilit\'es, Statistique et Modélisation, Universit\'e de Paris. \textsf{chassagneux@lpsm.paris}} \footnotetext[2]{%
Department of Mathematics, Imperial College London. \textsf{%
d.crisan@imperial.ac.uk}} \footnotetext[3]{%
Citigroup, London. \textsf{%
hchotai1@gmail.com}}

\renewcommand{\thefootnote}{\arabic{footnote}}

\vspace{2mm}

 \noindent decoupling
field, \noindent carbon markets, \noindent emission trading systems, \noindent market stability reserve, \noindent multiple compliance periods \vspace{2mm}

\noindent \textbf{MSC Classification (2020):} Primary 60H30; secondary 91G80.

\vspace{5mm}

\section{Introduction}
Carbon emission markets have been implemented in several regions worldwide as a measure to mitigate against climate change.    These are cap and trade schemes, where a regulator sets a cap on the total amount of emissions of all the participants in a particular market. The regulator releases a number of allowances, the number being equal to this aggregate cap. For each unit of emissions made, a regulated firm must surrender one allowance. The allowances can be traded amongst firms in the market, typically called a carbon market. Within a cap and trade scheme, firms that can reduce their emissions cheaply will do so, and these firms will sell excess allowances. On the other hand, firms that can not reduce their emissions cheaply will buy allowances to cover their emissions. If the aggregate cap is set appropriately, emissions reduction can be achieved.

The European Union (EU) has had its own emissions trading system (ETS) since 2005.
%
So far, there have been three different phases, with the fourth phase, phase 4, set to start in 2021. Every year, any operator that falls under the remit of the EU ETS must submit an emissions report outlining its level of emissions for the year and then surrender EUAs (European Union Allowances) for all of its emissions by April 30 of the following year. Each EUA is worth 1 tonne of CO\textsubscript{2}e, (equivalent tonnes of CO\textsubscript{2}). For each verified tonne of CO\textsubscript{2}e made by an operator and not accompanied by a EUA in the appropriate year, an operator must pay a penalty. Since 2013, the penalty has been set at \EUR{100} per tonne of CO\textsubscript{2}e, rising with the EU inflation rate \cite{euetshandbook}.
Phase 1 was a trial phase running from 2005 to 2008. During phase 1, only power generators and energy-intensive industries were covered. Almost all EUAs were given to firms for free. Towards the end of phase 1, it became clear that the total number of allowances issued would exceed the level of emissions. Since the EUAs from phase 1 could not be carried forward to the following phase, phase 2, the EUA price decreased to 0 towards the end of phase 1 \cite{phases1and2ecwebsite}.

Phase 2 ran from 2008 to 2012 and covered more sectors and companies than phase 1, and featured a lower cap on emissions. The proportion of EUAs allocated freely fell, and some EUAs were initially allocated through auctions. At the end of phase 2, firms could carry over, or `bank' any unused EUAs to the following phase, phase 3. This banking mechanism is expected to continue for transitions between all future phases and it is covered by the models in this paper.

As of 2020, phase 3 (2013-2020) is in operation. It covers more sectors and more greenhouse gases (GHGs) than previous phases. In addition, auctioning is now the default method for allocating allowances \cite{euetswebsite}. Since 2019, the EU ETS has also featured a market stability reserve (MSR). The aim of the MSR is to address the large surplus of allowances currently in the market. Each year, the European Commission will consider the total number of allowances in circulation. This number will, according to pre-defined rules, be used to decide whether a proportion of allowances that would have been released in the following year will be placed into the reserve, or whether allowances in the reserve will be released into the market. The reserve began with 900 million allowances that were deducted from auctioning volumes in the years 2014-2016 \cite{msrwebsite}. Full details of the MSR can be found in \cite{msrjournal}. The multi-period model presented in this paper is appropriate for modelling a MSR in a setting with two compliance periods and can be appropriately extended to model an MSR with an arbitrary number of compliance periods; see Example \ref{msr cap example} in the following section.

In 2013, 8715 million tonnes of CO\textsubscript{2}e worth of EU emissions allowances were traded. This was higher than in any year of phase 2. The majority of allowances, approximately 6000 million tonnes of CO\textsubscript{2}e worth in 2013, are traded on an exchange \cite{euetsfactsheet2}. There is evidence that the EU ETS has led to significant emissions abatement without a significant reduction in competitiveness. In addition, the EU ETS may have also led some operators to consider innovation activities that would lead to emissions reduction; see \cite{martin2015impact} for a review of various economic studies. 

Besides the EU ETS, there are many other carbon markets in operation worldwide, and the results and analysis in this paper should be equally applicable to these other markets. Notable examples include the California cap-and-trade program and South Korea's emissions trading scheme. 

China is planning to introduce  a national emissions trading scheme for greenhouse gas emissions in 2020. In preparation for this, in 2013, seven pilots were introduced in seven different regions in China. Since their inception up until 31 July 2015, over 57 million tons of carbon had been traded  under the pilots, and this quantity was valued at US\$308 million \cite{margolis_dudek_hove_2015, swartz2016china}. By 2018, pilot schemes had begun in the regions of Shenzhen, Shanghai, Beijing and Guangdong. It is set to be fully functional in 2020 and, when it is, China's emissions trading scheme will be the largest carbon market in the world, covering more CO\textsubscript{2}e of GHGs than any other such market. In 2019, the Energy Transitions Commission (ETC) released a report explaining that China can achieve net zero carbon emissions while becoming a fully developed economy by 2050 \cite{etc2019china}. 

In 2019, the World Bank stated that, in 2018, governments raised approximately US\$44 billion in carbon pricing revenues over 2018, constituting an increase of approximately US\$11 billion compared with the corresponding amount for 2017 \cite{ramstein2019state}. The High-Level Commission on Carbon Prices, in \cite{stern2017report}, concluded that achievement of the target within the Paris Agreement would be consistent with a carbon price level of between US\$40 and US\$80 by 2020 and between US\$50 and US\$100 by 2030. According to the World Bank's report, less than five percent of global emissions covered by carbon pricing initiatives are priced at this level; the majority are priced below this level. In 2019 a report convened by the Science Advisory Group to UN Climate Action Summit 2019 \cite{unitedinscience2019} concluded that countries' intended nationally determined contributions (NDCs) would roughly need to be tripled to be consistent with the target of at most a 2$^{\circ}$C rise in mean global temperature from pre-industrial levels.

For the EU ETS, between 2017 and 2018, total emissions from stationary installations declined by 4.1\%. Overall, total ETS emissions from stationary installations have declined by around 29\% between 2005 and 2018. It is expected that, stationary emissions are set to decrease by 36\% compared with 2005 levels by 2030 and that, even with additional measures, this reduction would be 41\%, which is still lower than the target value of 43\% \cite{euetstrendsprojections2019}. 

Carbon price formation is complex and can be approached in many different ways. This paper presents one particular approach. Broadly speaking, models for pricing emissions allowances in carbon markets can be separated into three categories: full equilibrium, risk-neutral and reduced form models; see \cite{howison2012risk} for further details. Consider a market consisting of firms producing goods which cause emissions which is connected to an emissions trading system comprising a liquid market for emissions allowances. In a full equilibrium model, one considers the interaction of individual firms in the market. Such models often lead to an optimization problem in which firms optimize their production of goods and number of emissions allowances. Some examples of full equilibrium models can be found in \cite{bossy2014nash,bossy2015game,carmona2009clean,carmona2009optimal,carmona2010market}. In a reduced form model, the coupling or interdependence between the allowance price and the level of emissions is not modelled explicitly. For example, we may assume that the level of cumulative emissions follows a standard process, such as geometric Brownian motion, independently of the allowance price. Such models are more tractable numerically compared with full equilibrium models. Some examples of reduced form models can be found in \cite{carmona2011risk,chesney2012endogenous,hitzemann2013empirical,nel2009carbon}. Finally, in a risk-neutral model, the price of an emissions allowance and the level of emissions are both modelled directly, without reference to individual firms. For these models, results are derived by using the tools and methods of risk-neutral pricing. Examples of risk-neutral models include the model described in \cite{cetin2009pricing} and all models which use FBSDEs to model a carbon market, which are described below. The model studied in this paper fits into the same risk-neutral framework. Other models that do not necessarily fit into only one of the aforementioned categories can be found in \cite{borovkov2011jump,cetin2009pricing,daskalakis2009modeling,hinz2010fair}.

The model studied here is based on a class of singular forward-backward stochastic differential equations (FBSDEs). These equations have three distinct features: their forward and backward component are coupled, the terminal condition for the backward component  is a
discontinuous function of the terminal value of the forward component, and
the forward dynamics are  degenerate. In particular, the equation satisfied by the cumulative emissions process has no volatility term. Such equations have already been used to model the evolution of the cumulative emissions and
price of an emissions allowance in a carbon market such as the EU ETS; see \cite{carmona2012valuation,carmona2013singular,fbsdespringerbrief,howison2012risk}.

The novelty in this work is that we introduce a model with multiple  trading periods. More precisely we consider a market consisting of participating firms whose activities cause emissions during the time interval $[0,T]$. For a positive integer $q$, representing the number of periods, the overall time interval $[0,T]$ is divided into $q$ trading periods:  $[T_0:=0, T_1]$, $[T_1,T_2]$,..., $[T_{q-1}, T_q:=T]$. During any trading period $[T_{k-1},T_k]$, for $1 \leq k \leq q$, emissions regulation is in effect. Let $(E_t^q)_{t \in [0,T]}$ be a real valued continuous process representing, at time $t$, the cumulative emissions made in the market up to time $t$. For every integer $0 \leq k \leq q$, at time $T_k$, the regulator records the level of cumulative emissions $E_{T_k}^q$, and, for each $0 \leq k \leq q-1$, a cap on the level of emissions at $T_{k+1}$ is defined. At each time $T_{k+1}$, the regulator checks whether the emissions made during the $[T_{k}, T_{k+1}]$ have exceeded the time $T_{k+1}$ cap and market participants must pay a penalty for each unit of emissions above the cap. We will denote by $(Y^{q}_t)_{t \in [0,T]}$  the spot price of an allowance certificate at time $t$. At each time $T_{k+1}$ the spot price will depend on the value of the cumulative emissions $E_{T_{k+1}}$ but also on the cap imposed by the regulator, which in turn, will depend on $E_{T_{k}}$.        

This model is more realistic than the single-period model introduced in previous work \cite{carmona2012valuation, carmona2013singular, fbsdespringerbrief}. In the single trading period model, any unused allowances become worthless at the end of the trading period. The multi-period model allows for the caps on the level of emissions $E^{q}_{T_{k+1}}$ at time  ${T_{k+1}}$ to depend on the level of emissions $E^{q}_{T_{k}}$ accumulated from time 0 up to time $T_k$, the beginning of the $k$-th trading period. In this framework, unused allowances can be ported to the next period (except for the final period) depending on the mechanism being modelled. This permits the modelling of mechanisms including the banking, borrowing or withdrawal of allowances between compliance periods. See Example \ref{simple cap example} below for further details.

The main result of the paper is to give a characterization of the pair of process $(E^{q}_t, Y^{q}_t )_{t \in [0,T]}$ as the unique solution of a set of $q$ FBSDEs that are linked through their transition values at times $T_k$, $k=0,\ldots,q-1$ and terminal conditions at times $T_k$, $k=1,\ldots,q$. In addition, the consecutive FBSDEs can be linked so as to model banking, borrowing and withdrawal of allowances, as described above. The linking of the FBSDEs in the multi-period model means that it is not possible to directly consider each FBSDE as a separate single-period model. This is the main technical difficulty of studying the multi period model. As usual, the study of this FBSDE is closely linked to the study of the associated value function (known as decoupling field in the FBSDE literature). This decoupling field can be considered to be an entropy solution to a degenerate quasilinear elliptic PDE. Even though we rely on  some important results given in \cite{carmona2013singularconservation}, we also establish new estimates concerning this value function, see e.g. Lemma \ref{le l1 estimate one period}, Lemma \ref{le control lipschitz} and Lemma \ref{le other control lipschitz} below. They constitute  a step forward in the study of singular FBSDEs and their associated decoupling fields.
    
As described above, a multi-period model is more applicable than a single period model because it allows one to model multiple compliance periods. One disadvantage of the multi-period model studied here, however, is that, for a $q$ period model, one must specify the end date $T_q$. This is important because, at $T_q$, the spot price $Y^{q}_{T_q}$ of an allowance certificate  is different from the one specified at every prior time $T_k$, for $1 \leq k \leq q-1$. The time $T_q$ is the time at which all emissions regulation ceases and this is why  $Y^{q}_{T_q}=0$ if at the time $T_q$ cumulative emissions are below the time $T_q$ cap. For a more realistic model, one can consider a model for a carbon market with no specified end date. In the setting of the EU ETS, for example, there is currently no time at which one can say with certainty that emissions regulation will cease or the banking of allowances to next period will be prohibited. In the second part of the paper, we introduce a model for a carbon market in operation over the time period $[0, \infty)$ with no end date and show that it is well posed under certain conditions. More precisely, we give a characterization of a pair of processes, $(E^{\infty}_t, Y^{\infty}_t )_{t \in [0,\infty)}$, say, $E^{\infty}$ representing the cumulative emissions and $Y^{\infty}$ the allowance price, as the unique solution of an  infinite sequence of FBSDEs that are linked through their transition values at times $T_k$, 
$k\ge 1$. Moreover we show that, under reasonable conditions, the spot price $Y^{q}_{t}$ of an allowance certificate for the $q$-period model 
converges, as the number of periods $q$ increases, to the spot price 
$Y^{\infty}_{t}$ of an allowance certificate for the infinite period model. Again, the results are obtained by a careful study of the associated decoupling field.

Carbon markets face many criticisms. Some critics claim that they reduce industries' competitiveness, while others believe that the average carbon price today is not high enough to motivate a substantial reduction in greenhouse gas emissions. Proponents of emissions trading systems claim that they lead to real emissions reductions when regulators operate them in an appropriate way. In any case, it is clear that emissions trading systems are becoming increasingly important and prevalent. Scientific, particularly mathematical, studies of them are needed in order to expose more about their advantages, their shortcomings, and their efficient implementation. 
It is hoped that the results of this paper will improve the understanding of carbon markets and help regulators to implement them in a way that brings the greatest social benefit in the action against the effects of climate change.

The rest of this paper is organised as follows. In Section \ref{preliminaries}, we introduce the main assumptions and key notions used in using singular FBSDEs to model carbon markets in a multi-period setting, and give generic statements of the main results. In particular, we state the wellposedness of a class of singular FBSDEs with the terminal condition for the backward equation that is a discontinuous function of the terminal value of the forward equation, and the forward dynamics may be degenerate (having no volatility term). This FBSDE is autonomous in the sense that it does not depend on other processes. This property holds for all of the $q$ FBSDEs comprising a $q$ period model. In Section \ref{mainresults1} we state and prove the wellposedness of the multi-period model. We also present several new results concerning singular FBSDEs that are needed to study the infinite period model. In Section \ref{mainresults2}, we state and prove the wellposedness of the infinite period model as well as the convergence of the value function for a $q$-period multi-period model to a value function for the infinite period model, as $q$ tends to infinity.  The paper is concluded with an appendix in which some results for single period models that were used in this paper are presented.

\subsubsection*{Notation.}

In the  following we will  use the following spaces  
\begin{itemize}
\item 
For  fixed $0\le a<b<+\infty$ and $I=[a,b]$ or $I=[a,b)$, { $\mathcal{S}^{2,k}(I)$ is the
set of  $\mathop{}\!\mathbb{R}^k$-valued c\`adl\`ag\footnote{French acronym for right continuous with left limits.} $\mathcal{F}_t$-adapted processes $Y$, s.t.\\[-4mm]  
\begin{align*}
\|Y\|_{\mathcal{S}^2}^2:= \mathbb{E}\left[{\sup_{t \in I} |Y_t|^2 } %
\right] <\infty .
\end{align*}
Note that we may omit the dimension and the terminal date in the norm notation as this will be clear from the context. $\mathcal{S}^{2,k}_\mathrm{c}(I)$ is the subspace of process with continuous sample paths.} 
\\
We also consider ${\mathcal{S}}^{2,k}( [0,\infty))$ the vector space of c\`adl\`ag adapted processes $Y$, with values in $\R^k$, and such that $\mathbb{E}\left[\sup_{0 \leq t  \leq b} |Y_t|^2  \right] <\infty $ for every $b > 0$.  ${\mathcal{S}}^{2,k}_{\mathrm{c}}( [0,\infty))$ denotes the subspace of such processes having continuous paths.

\item For fixed $0\le a<b<+\infty$, and $I=[a,b]$ or $I=[a,b)$ again, we denote by { $\mathcal{H}^{2,k}(I)$ the set
of $\mathop{}\!\mathbb{R}^k$-valued progressively measurable processes $Z$, such that \\[-4mm] 
\begin{equation*}
\|Z\|_{\mathcal{H}^2}^2 := \mathbb{E} \left[{\int_I |Z_t|^2 dt} \right] 
<\infty.  
\end{equation*}%
$\mathcal{H}^{2,k}([0,\infty))$ is the set
of $\mathop{}\!\mathbb{R}^k$-valued progressively measurable processes $Z$, such that 
$ \mathbb{E} \left[{\int_0^b |Z_t|^2 dt} \right] 
<\infty$, for all $b>0$.
}
\end{itemize}

\noindent  For any process $X$ and time $t$, we denote by $X_{t^{-}}$ and $X_{t^{+}}$ respectively, the left and right limit of the process $X$ at $t$, i.e  
\[
X_{t^{-}} = \lim_{s \uparrow t} X_s, \  \ \
X_{t^{+}} = \lim_{s \downarrow t} X_s\,.
\]
%

\noindent For $\phi:\R^d \times \R \rightarrow \R$, measurable and non-decreasing in its second variable,
the functions $\phi_{-}$ and $\phi_{+}$ are the left and right continuous versions, respectively defined, for $(p,e) \in \mathbb{R}^{d} \times \mathbb{R}$, by,
	\begin{align}
	\begin{aligned}
	\label{phi minus plus definition}
	\phi_{-}(p,e) &= \sup_{e^{\prime} < e} \phi(p, e^{\prime}) \\
	\phi_{+}(p,e) &= \inf_{e^{\prime} > e} \phi(p, e^{\prime}).
	\end{aligned}
	\end{align}
Moreover, we denote by $\| \cdot \|_\infty$ the essential supremum:
\begin{align*}
\| \phi \|_\infty = \mathrm{esssup}_{(p,e)\in \R^d \times \R} |\phi(p,e)| \;.
\end{align*}


\section{Framework and main result}
\label{preliminaries}
\subsection{Framework for the multi-period model}
Let  $( \Omega, \mathcal{F}, \mathbb{P})$ be a complete probability space.
We denote by $W$ a $d$ dimensional Brownian motion defined on $( \Omega, \mathcal{F}, \mathbb{P})$ started at $t=0$, and $\{\mathcal{F}_t\}_{0 \leq t \leq T}$ the complete filtration generated by the Brownian motion $W$. In the following, we consider a market with $q \geq 2$ trading periods, denoted $[T_0, T_1]$, $%
[T_1, T_2]$,...,$[T_{q-1},T_{q}]$ with $T_0:=0<\dots<T_k<\dots<T_q=:T$ for $T>0$. The market is governed by  three processes $(P,E,Y)$ all defined on  $( \Omega, \mathcal{F}, \mathbb{P}):$  \\
\begin{itemize}
\item The process $(Y_t)_{0 \leq t \leq T}$ represent the \emph{spot price}\footnote{Although European Union Allowance (EAU) futures are most commonly traded in the EU ETS, we model the spot price rather than the futures price to  simplify the presentation in the multi-period model. For a one period model, futures prices can be directly modelled. Note that, in our setting of a constant, deterministic interest rate $r$, the spot and futures prices only differ by a multiplicative deterministic discount factor.} of a carbon emissions allowance. The constant $r$
will denote the instantaneous risk-free interest rate, which will be assumed to be fixed and deterministic throughout this paper. $r$ is such that investment of $x_0$ at time $0$ yields $e^{rt}x_0$ at time $t$, for any $t \in [0,T]$. We will assume that the allowances are traded assets, and that the discounted price process $(e^{-rt}Y_t)_{t \in [T_k,T_{k+1})}$ is an $\mathcal{F}_t$-adapted martingale (see also Remark \ref{risk neutral remark}). 
Equivalently, the dynamics of $Y$ can be written
\begin{align}
        \label{discounted A dynamics}
        Y_{t_2} = Y_{t_1} + \int_{t_1}^{t_2} rY_{s} ds + \int_{t_1}^{t_2} Z_s d W_s,
        \end{align}
        for every $T_k \le t_1 \le t_2 <T_{k+1}$, where $(Z_t)_{t\in [T_{k},T_{k+1})}$ is a progressively measurable process such that $\mathbb{E} \bigg[ \int_{T_{k}}^{T_{k+1}} |{Z_s}|^2 d s \bigg] < \infty$, $k \le q$. 
\item The process $(P_t)_{0 \leq t \leq T}$
represents factors in the market that will also drive emissions. We assume that this process is purely autonomous and independent of the level of cumulative emissions and the allowance price. 
For example, in the presentation of \cite{carmona2012valuation} for an electricity market
with emissions regulation, $(P_t)_{0 \leq t \leq T}$ could be a vector
consisting of fuel prices and an inelastic demand curve for electricity. Setting $P_{0} = p$, a deterministic constant, we assume
that its dynamics are given by 
\begin{align}  \label{multi period P equation}
\mathop{}\!\mathrm{d} P_t = b(P_t) \mathop{}\!\mathrm{d} t + \sigma(P_t) %
\mathop{}\!\mathrm{d} W_t, \quad t \in [0, T], 
\end{align}
for functions $b$ and $\sigma$ such that strong existence and uniqueness
holds for the SDE \eqref{multi period P equation}. (The conditions on all
coefficient functions will be made precise below and in the following
section).

\item The process  $(E_t)_{0 \leq t \leq T}$ represents the cumulative emissions in the market; it results from integrating the quantity $\mu(P_t,Y_t)$, 
where $\mu: \mathbb{R}^d
\times \mathbb{R} \rightarrow \mathbb{R}$ is a function representing the
market emissions rate. 
In other words, we assume
that the dynamics of the cumulative its dynamics are given by 
\begin{align}  \label{multi period E equation}
\mathop{}\!\mathrm{d} E_t = \mu(P_t, Y_t)
\mathop{}\!\mathrm{d} t,\ \ \ t\in [0,T], 
\end{align}
\end{itemize} 
\begin{remark}
\label{risk neutral remark}
We are implicitly assuming here that we work under a risk neutral probability measure $\P$. In this setting, the discounted future cash flow of any tradable asset is a martingale.
\end{remark}

\noindent To summarise, we shall assume that the 4-tuple $(P_t, E_t, Y_t, Z_t)_{t_0 \leq t \leq T}$ satisfies the following forward-backward stochastic differential equation on each period $[T_k,T_{k+1})$, $k \le q$: 

\begin{align}  \label{general single period fbsde without p}
\begin{aligned} \mathop{}\!\mathrm{d} P_t &= b(P_t) \mathop{}\!\mathrm{d}
t + \sigma(P_t) \mathop{}\!\mathrm{d} W_t, & 
\\ \mathop{}\!\mathrm{d} E_t &= \mu(P_t, Y_t) \mathop{}\!\mathrm{d} t, &
\\ \mathop{}\!\mathrm{d} Y_t &= r Y_t
\mathop{}\!\mathrm{d} t + Z_{t} \mathop{}\!\mathrm{d} {W_t}. \end{aligned}
\end{align}

This description is obviously incomplete as one must understand what happens at the end of each period in other to link them together and hopefully obtain a market process on the whole time interval $[0,T]$. We now describe the mechanisms that are put in place in the ETS market and that we will take into account in our model.

\noindent For every integer 
$0 \leq k \leq q - 1$, the number of allowances in circulation at $T_{k-1}$,
the start of the $[T_{k-1}, T_k]$ period, will be assumed to be a
deterministic function of $E_{T_{k-1}}$, namely $\Gamma_k(E_{T_{k-1}})$ 
where $ \R \ni \mathfrak{e} \rightarrow \Gamma_k(\mathfrak{e}) \in \R$. 
This is the cap on emissions during that period;
compliance occurs at time $T_k$, if and only if $E_{T_{k}} - E_{T_{k-1}} <
\Gamma_k(E_{T_{k-1}})$ or equivalently if $E_{T_{k}} < {\Lambda}_k
(E_{T_{k-1}})$, where ${\Lambda}_k (\mathfrak{e}) := \Gamma_k(\mathfrak{e}) + \mathfrak{e}$. 
The quantity ${\Lambda}%
_k(E_{T_{k-1}})$ represents then the cap on cumulative emissions at $T_k$, 
 namely a cap for
all emissions from time $0$ to time $T_k$. 
For every $k
\geq 1$, at time $T_k$ there is a penalty for non-compliance $\rho_k$, which is usually set to be equal to 1, incurred if the
cumulative emissions up to that time $E_{T_k}$ have exceeded the cap ${\Lambda}_k$. 
In the sequel, we will work with penalty functions $\Lambda$ which belong to the  following class.
\begin{Definition}
\label{cap functions set definition}
Let $\Theta$ be the class of functions $\Lambda : \R \rightarrow \R$ such that
$\mathfrak{e} \mapsto \Gamma(\mathfrak{e}) := \Lambda(\mathfrak{e}) - \mathfrak{e}$ is monotone decreasing
and satisfies $%
\lim_{\mathfrak{e} \rightarrow +\infty} \Gamma(\mathfrak{e}) = -\infty$.
\end{Definition}

%

\noindent We now give examples of cap functions that can be encountered in practice.
\begin{example}\label{simple cap example} 
We  give examples to show how the cap functions $({\Lambda}_k)_{1 \leq k \leq q}$ can be chosen to model different mechanisms that are in force in the EU ETS market, namely
\begin{itemize}
        \item{\emph{Banking:} allowances that are not used in one period can be carried forward for compliance in the next period.}
        \item{\emph{Withdrawal:} for any $1 \leq i \leq q-1$, if the cap on emissions is exceeded at $T_i$, then the regulator removes a quantity of allowances from the $[T_{i}, T_{i+1}]$ market allocation. The quantity of allowances removed is equal to the level of excess emissions at $T_i$.}
        \item{\emph{Borrowing:} for any $1 \leq i \leq q-1$, firms may trade some of the allowances to be released at $T_i$ during $[T_{i-1},T_i]$. If each trading period represents a year, this means that firms can, in a particular year that is not the final year, use the following year's allowance allocation for compliance.}
\end{itemize}

Suppose that the regulator
releases $c_{k+1} \ge 0$ allowances into circulation at each time $T_k$ for $%
k=0,1,...,q-1$.


\noindent To take into account banking, borrowing and withdrawal, we can set
                                \begin{align}
                                \label{banking borrowing withdrawal cap}
                                \Gamma_k(\mathfrak{e}) = \sum_{i=1}^{ (k+1) \wedge q  } c_i - \mathfrak{e},
                                \end{align}
For banking and withdrawal only, we can set
                                \begin{align}
                        \Gamma_k(\mathfrak{e}) = \sum_{i=1}^{ k  } c_i - \mathfrak{e},
                        \end{align}
                        $\left.\right.$\\[-8mm]
                for every $1 \leq k \leq q$.    
\end{example}

\noindent We can also present a very simple example in which the functions ${\Lambda}_k$ are not all constant functions.
\begin{example}
	\label{msr cap example}
	[Simple market stability reserve for a two period model]
	Let $q=2$ for a two period model. Suppose that, similarly to Example \ref{simple cap example}, for each $k =0, 1$, the regulator has a quantity of allowances $c_{k+1} \geq 0$ to be released into the market at $T_{k}$ and to be used for compliance at any time after $T_{k}$ To specify the mechanism that links the two periods, we simply need to specify the two functions $\Gamma_1$ and $\Gamma_2$ or, equivalently, ${\Lambda}_1$ and ${\Lambda}_2$. Let ${\Lambda}_1 = c_1$, thus $\mathfrak{e} \mapsto \Gamma_1(\mathfrak{e}) = c_1 - \mathfrak{e}$. Assuming  $E_0 = 0$, the cap at $T_1$ is simply a constant, equal to $c_1$.
	
	Suppose that, at $T_1$, the regulator considers the number of allowances in circulation instantaneously after all allowances for emissions up to $T_1$ have taken place and checks whether it is within the interval $[\kappa^{\mathrm{L}}, \kappa^{\mathrm{U}}]$ where $\kappa^{\mathrm{L}} < \kappa^{\mathrm{U}}$ and $\kappa^{\mathrm{L}}$ and are $\kappa^{\mathrm{U}}$ are, respectively, lower and upper thresholds. The regulator then adjusts the number of allowances in circulation in the following way. The regulator adds a fixed quantity $c$ if the number of allowances in circulation would be below the threshold, reduces the number of allowances in circulation by a proportion $(1-\alpha)$ of the total if the number of allowances in circulation would be above the threshold, and does not make an adjustment if the number of allowances would be within the threshold. Here, $c > 0$ and $0 < \alpha < 1$ are constants. Mathematically, this is expressed as
	\begin{align}
	\Gamma_2 (E_{T_2}) = 
	\begin{dcases}
	\tilde{\Gamma}_2(E_{T_1}) + c, & \text{if } \tilde{\Gamma}_2(E_{T_1}) < \kappa^{\mathrm{L}}, \\
	\tilde{\Gamma}_2(E_{T_1}), & \text{if } \kappa^{\mathrm{L}} \leq \tilde{\Gamma}_2(E_{T_1}) \leq \kappa^{\mathrm{U}},
	\\
	\alpha \tilde{\Gamma}_2(E_{T_1}), & \text{if } \tilde{\Gamma}_2(E_{T_1}) > \kappa^{\mathrm{U}},
	\end{dcases}
	\end{align}
	where $\tilde{\Gamma}_2(E_{T_1})$ represents the number of allowances in circulation at $T_1$ instantaneously before adjustment with $\tilde{\Gamma}_2$ given by
	\begin{align}
	\mathfrak{e} \mapsto \tilde{\Gamma}_2(\mathfrak{e}) = c_1 + c_2 - \mathfrak{e}.
	\end{align}
	
\noindent The above setting is a simple version of a market stability reserve. Such a reserve was established and began operation in January 2019 in the EU ETS (\cite{msrjournal}). This reserve is designed to reduce the large surplus of allowances that has built up in the EU ETS since phase 2. In the EU ETS, if the total number of allowances in circulation is too high, then 12\% of these allowances will be removed and placed in the reserve. Similarly, if the number of allowances in circulation is too low (less than 400 million) in a given year, then a quantity of allowances up to 100 million will be released from the reserve and added to the volume of allowances set to be released through auction. See \cite{msrjournal} for full details.
	
To model a market stability reserve more realistically for a $q$ period model, where $q > 2$, we would need to study a model in which, for each $2 < k \leq q$, the cap on emissions at $T_k$ depends not only on $E_{T_{k-1}}$, but on all quantities $E_{T_1}$, $E_{T_2}$,..., $E_{T_{k-1}}$. Although we do not consider such a model in this paper, the model presented here can be extended to such a setting and, under additional conditions, the results regarding existence and uniqueness can be proved.
\end{example}

\begin{remark}
The modelling assumption of the emissions process being continuous at all
times including the compliance times can be justified in the following way: Assuming continuity
at each compliance time is required to specify the initial and terminal
conditions for those FBSDEs, and it implies that the solutions of the
different FBSDEs defining the multi period model are not independent. This
is realistic and means that a multi period model is different to several
separate copies of a single period model. In a realistic market setting, an
increase or decrease in emissions requires an adjustment in the factors of
production and we argue that this can not be carried out instantaneously in
such a way that the emissions process would develop a point of discontinuity
at any time. Another argument is the following. For each $1 \leq k \leq q$,
the cap at time $T_{k}$, which will affect the market dynamics over $%
[T_{k},T_{k+1}]$ is a deterministic function of $E_{T_{k-1}}$; it is already
known at time $T_{k-1}$. Based on this, it is reasonable to assume that $E$
will not have a jump or any kind of discontinuity at $T_k$ for any $1 \leq k
\leq q$.
\end{remark}

We now describe the link between periods for the $Y$-process, it follows from the following heuristics\footnote{\textcolor{black}{See e.g. \cite{carmona2010market} for an equilibrium argument in the one period setting.}}: If the cumulative emissions at time $T_k$ is less than the cap at that time, i.e., 
$E_{T_{k}}< \Lambda_k(E_{T_{k-1}})$, then the spot price of the allowance will be ported to the next period, in other words, 
$\lim_{t\rightarrow T_k}Y_t=Y_{T_k}$. However, if $E_{T_{k}} > \Lambda_k(E_{T_{k-1}})$,
then the spot price should converge to the penalty for non-compliance, in other words, $\lim_{t\rightarrow T_k}Y_t=1$.
For the last period, as the market ends at $T_q$, we set by convention $Y_{T_q} = 0$.

%
%

\vspace{4mm}
Before stating our main result concerning the existence  and uniqueness of an equilibrium for the market described above in this multi-period setting, we give the assumptions on the coefficient functions parameters that will be used throughout this paper.



\begin{Assumption}\label{ass coef}
\label{delarue assumptions}  The functions $b:\mathbb{R}^d
\rightarrow \mathbb{R}^d$, $\sigma:\mathbb{R}^d \rightarrow \mathbb{R}^{d
\times d}$ and $\mu: \mathbb{R}^d \times \mathbb{R} \rightarrow \mathbb{R}$
are such that there exist three constants $L \geq 1$, $l_1, l_2 >0$, $1/L
\leq l_1 \leq l_2 \leq L$ satisfying 

\begin{enumerate}
\item $b$ and $\sigma$ 
are $L$-Lipschitz continuous:  
\begin{align}
|b(p) - b(p^{\prime})| + |\sigma(p) - \sigma(p^{\prime})| \leq L |p -
p^{\prime}|, \quad p,p^{\prime} \in \mathbb{R}^d.
\end{align}

\item {$\mu$ 
is $L$-Lipschitz continuous, satisfying  
\begin{align}
|\mu(p,y) - \mu(p^{\prime},y^{\prime})| \leq L\left(|p - p^{\prime}| + |y -
y^{\prime}|\right), \quad p,p^{\prime} \in \mathbb{R}^d, y,y^{\prime} \in 
\mathbb{R}.
\end{align}
Moreover, for any $p \in \mathbb{R}^d$, the real function $y \mapsto \mu(p,y)$
is strictly decreasing and $\mu$ satisfies the following monotonicity
condition  
\begin{align}  \label{monotonicity technical assumption}
l_1 |y - y^{\prime}|^2 \leq (y - y^{\prime}) \left(\mu(p, y^{\prime}) -
\mu(p, y)\right) \leq l_2 |y - y^{\prime}|^2, \quad p \in \mathbb{R}^d,
y,y^{\prime} \in \mathbb{R}.
\end{align}%
} 
\end{enumerate}
\end{Assumption}

\begin{Remark} 
The strict monotonicity of $\mu$  \eqref{monotonicity technical assumption} is required to prove some bounds used in the analysis of singular FBSDEs such as \eqref{general single period fbsde without p}. In practical applications, $\mu$ can be interpreted as an emissions rate function. In such settings, at a time $t$, given an allowance price $Y_t$ and a vector of factors that drive emissions, $P_t$, $\mu(P_t, Y_t)$ is the rate of emissions per unit time. So the strict monotonicity amounts to assuming that the emissions rate is strictly decreasing in the allowance price. This is economically reasonable because higher allowance prices can be expected to promote lower level of emissions. In the literature, when \eqref{general single period fbsde without p} has been studied for a carbon market operating within an electricity market, $\mu$ is explicitly constructed, using functions that arise in the modelling of the electricity market, in such a way that guarantees that the strict monotonicity of $\mu$ holds \cite{carmona2012valuation,fbsdespringerbrief,howison2012risk}. Also, as described in \cite{carmona2013singularconservation}, we can interpret the strict monotonicity of $\mu$ as convexity of the anti-derivative of $-\mu$, which is standard when studying scalar conservation laws. The FBSDE \eqref{general single period fbsde without p} has strong links to the theory of scalar conservation laws: as shown in \cite{carmona2013singularconservation}, Section 3, the value function constructed for \eqref{general single period fbsde without p} can be related to a solution of the inviscid Burgers equation.
\end{Remark}

\noindent Our first main result is the well-posedness for the multi-period pricing problem.

\begin{Theorem}
        \label{multi period model main theorem}
\textcolor{black}{Let $q \geq 1$ be an integer representing the number of trading periods and let $\Lambda_k \in \Theta$, $1 \le k \le q$ be the cap functions, recall Definition \ref{cap functions set definition}. We consider a $q$ period multi-period model over $[T_0, T_1]$, ...., $[T_{q-1},T_q]$ with $T_0 := 0$ and $T_q := T$.} Under Assumption \ref{ass coef},
there exists a unique c\`{a}dl\`{a}g process $(Y_t)_{0 \leq t \leq T} \in \mathcal{S}^{2,1}([0,T])$, a continuous process $(E_t)_{0 \leq t \leq T} \in \mathcal{S}^{2,1}_{c}([0,T])$, a continuous process $(P_t)_{0 \leq t \leq T} \in \mathcal{S}^{2,d}_{c}([0,T])$ and a process $(Z_t)_{0 \leq t \leq T} \in \mathcal{H}^{2,d}([0,T])$  satisfying the dynamics \eqref{general single period fbsde without p} on each period $[T_{k-1},T_k)$, $1 \leq k \leq q$. The process $Y$ is  continuous on $[T_{k-1},T_k)$; it can have a jump at $T_k$. There it satisfies, for every $1 \leq k \leq q$, almost surely,
        \begin{align}
        \begin{aligned}
        \label{multi period relaxed terminal condition in theorem}
         Y_{{T_k}^{-}} &= Y_{T_k},  && \text{if } E_{T_k} < {\Lambda}_k (E_{T_{k-1}}), \\
        Y_{{T_k}^{-}} &= 1, && \text{if } E_{T_k} > {\Lambda}_k (E_{T_{k-1}}), \\
        Y_{T_k} \leq & Y_{{T_k}^{-}} \leq 1, && \text{if } E_{T_k} = {\Lambda}_k (E_{T_{k-1}}), 
        \end{aligned}
        \end{align}
with, by convention, $Y_{T_q} = 0$.
        
\noindent Moreover, there exists a unique function $v$
$$[0,T] \times \mathbb{R}^{d} \times \mathbb{R}\times \mathbb{R}  \ni (t,p,e,\mathfrak{e}) \mapsto v(t,p,e,\mathfrak{e}) \in \R $$
such that, for each $\mathfrak{e} \in \R$, $(t,p,e)\mapsto v(t,p,e,\mathfrak{e})$ is continuous on $[T_{k-1},T_k)\times \mathbb{R}^{d} \times \mathbb{R}$,  and $Y_t = v(t,P_t,E_t,E_{T_{k-1}})$, $T_{k-1} \le t <T_k$, $1 \le k \le q$.   By convention, $v(T,p,e,\mathfrak{e})=0$, for $(p,e,\mathfrak{e}) \in \mathbb{R}^{d} \times \mathbb{R}\times \mathbb{R} $. 
\end{Theorem}

\begin{Remark} \label{re th multi-period}
\begin{enumerate}[i)]
\item The function $v$ is a key object to obtain existence and uniqueness to the FBSDEs and is known as the \emph{decoupling field} for equation \eqref{general single period fbsde without p}. In our setting, it appears also as a pricing function for the allowance contract, with respect to the underlying process $(P,E)$. Further properties of the function $v$ are given in Proposition \ref{pr pricing function multi-period} below.
\item Here, we consider a \emph{càdlàg} $Y.$  This differs slightly from what is done in the one period setting \cite{carmona2013singularconservation} where the \emph{càglàd} version is considered. This is merely a convention that simplifies our presentation in the multi-period setting.
\item For a single period model comprising a single copy of equation \eqref{general single period fbsde without p} on a time interval $[0,T]$ say, it is shown in \cite{carmona2013singularconservation} that in general, given a monotone increasing real valued function $\phi$ taking values in $[0,1]$ (note that $\phi$ is not assumed to be Lipschitz continuous), one can not construct a process $Y$ satisfying the dynamics \eqref{general single period fbsde without p} and such that $Y_T = \phi(E_T)$ almost surely. It is shown there that existence and uniqueness of solutions to the FBSDE does hold when, instead, the relaxed terminal condition,
\begin{align}
\label{relaxedterminalcondition general}
\mathbb{P}[ \phi_{-}(E_T) \leq Y_{T^{-}}  \leq \phi_{+}(E_T)] = 1,
\end{align}
is imposed. To model a single period carbon market in which the penalty for over emission is 1 and the cap on emissions is $\lambda$, one would typically take $\phi(x) = \mathbf{1}_{[\lambda, +\infty)}(x)$ for every $x \in \R$. The terminal condition, \eqref{multi period relaxed terminal condition in theorem}, presented in Theorem \ref{multi period model main theorem}, is the multi period version of this terminal condition. Notice that in the multi-period model, the cap on emissions at time $T_k$ is $\Lambda_k(E_{T_{k-1}})$.
\end{enumerate}
\end{Remark}

\subsection{Infinite period model}

The  multi period model introduced in the previous section is more realistic and applicable than a single period model because it allows one to model multiple times at which compliance occurs and a new allowance allocation is released into a carbon market. One disadvantage of this, however, is that, for a $q$ period model, one must specify the end date $T_q$. 
The time $T_q$ is the time at which all emissions regulation ceases and this is why the terminal condition at this time specifies that allowances at $T_q$ will have price $Y_{T_q}=0$ if the time $T_q$ cumulative emissions are below the time $T_q$ cap.

For an even more realistic model, one might prefer to consider a model for a carbon market with no specified end date. Phase 2 of the EU ETS was followed by Phase 3, with no interruptions. Although new rules came into effect at the start of Phase 3, any Phase 2 allowances could be carried forward (banked) into Phase 3 and used for compliance later and, according to the EU ETS handbook \cite{euetshandbook}, this can be expected to continue for Phase 4 and all future phases. Therefore, in the setting of the EU ETS, there is currently no time at which one can say that emissions regulation will cease or the banking of allowances will be prohibited.

We thus introduce and study a model with an infinite number of period to address the aforementioned limitations. This can be thought of as a model for a carbon market in operation over the time period $[0, \infty)$ with no end date. The cap and trade periods are still connected through the \emph{banking, withdrawal and borrowing} rules.
%
For this infinite period model, we shall work under the following setting:
\begin{Assumption}\label{ass cap infinite model}
We set $T_k = k \tau$, for every $k \geq 0$, where $\tau > 0$. The cap functions $({\Lambda}_k)_{k \geq 1}$ are constants and given by, 
$
\Lambda_k = k \lambda
$, for $k \ge 1$ and $\lambda > 0$. (By a slight abuse of notation, we make no  difference between the function and its constant value.)
\end{Assumption}


%
\noindent  
We shall also require one of the two following assumptions:
\begin{Assumption}\label{ass infinite model coef}
The coefficient $b$ and $\sigma$ are such that for $p \in \R^d$, $b_i(p)$ and $\sigma_{i\cdot}(p)$ only depend on $p_i$. Moreover, there exists $\beta \in \R$ such that, denoting $L_{b}$ the Lipschitz constant of $b$,
\begin{align*}
r - L_{b} \ge \beta > 0\;.
\end{align*}
\end{Assumption}
\begin{Assumption} \label{ass unif ellip}
\begin{enumerate}[i)]
\item The matrix $\sigma$ is uniformly elliptic, namely, there exists $\beta > 0$ such that
\begin{align*}
\upsilon^\top \sigma(p)\sigma(p)^\top \upsilon \ge \beta |\upsilon|^2\;,\quad \forall (p,\upsilon) \in \R^d\times\R^d\;.
\end{align*}
\item The interest rate $r$ is strictly positive.
\end{enumerate}
\end{Assumption}

\begin{Remark} 
\begin{enumerate}[i)]
\item The constant $\tau > 0$  represents the length of each period, which is typically one year. In this case, the constant $\lambda$ is the yearly cap.
\\
In light of Example \ref{simple cap example}, assuming the $\Lambda_k$ constant is not much of a restriction in applications.
Assumption \ref{ass infinite model coef} (resp. Assumption \ref{ass unif ellip}(i)) is a technical assumption used to guarantee that some Lipschitz constant does not explode when considering an infinite number of periods, see  Lemma \ref{le control lipschitz} (resp. Lemma \ref{le other control lipschitz}) below.
\item Assumption \ref{ass infinite model coef} implies, in particular, that the rate $r$ is strictly positive. 
\end{enumerate}

\end{Remark}

\noindent Our main result in this setting is the following.
\begin{Theorem}\label{th infinite model 1}
Let Assumptions \ref{ass coef} - \ref{ass cap infinite model} - \ref{ass infinite model coef} or \ref{ass coef} - \ref{ass cap infinite model} - \ref{ass unif ellip} hold. 
Then there exist processes $P \in \cS^{2,d}_c([0,\infty))$, $E \in \cS^{2,1}_c([0,\infty))$, $Y \in \cS^{2,1}([0,\infty))$ and $Z \in \cH^{2,d}([0,\infty))$ satisfying on each period $[T_k,T_{k+1})$:
\begin{align}
\label{fbsde time homogeneous}
\begin{aligned}
\dif P_t &= b(P_t) \dif t + \sigma(P_t) \dif W_t  \\ 
\dif E_t &= \mu(P_t, Y_t) \dif t  \\ 
\dif Y_t &= rY_t \dif t + Z_{t} \dif {W_t}.
\end{aligned}
\end{align}
The process $Y$ is continuous on $[T_{k-1},T_k)$. It can have a jump at $T_k$, $1 \le k$, where it satisfies, almost surely
        \begin{align}
        \begin{aligned}
        \label{infinite period relaxed terminal condition in theorem}
        Y_{{T_k}^{-}} &= Y_{T_k}, && \text{if } E_{T_k} < {\Lambda}_k , \\
        Y_{{T_k}^{-}} &= 1, && \text{if } E_{T_k} > {\Lambda}_k, \\
        Y_{T_k} \leq & Y_{{T_k}^{-}} \leq 1, && \text{if } E_{T_k} = {\Lambda}_k.      
        \end{aligned}   
        \end{align}
  Moreover, there exists a continuous function 
  $w:[0,\tau)\times \R^d \times \R \rightarrow \R$
  such that
$ Y_t = w(t-T_{k-1},P_t,E_t-\Lambda_{k-1})$, $t \in [T_{k-1},T_{k})$,  $k \ge 1$. Setting, for  $e \in \R$,
\begin{align}
\label{Theta k definition in introduction}
\Phi(p,e) = \begin{dcases} w(0, p, e - \lambda), & \text{if }       e < \lambda, \\
1, & \text{otherwise.}
\end{dcases}
\end{align}
the function $w$ satisfies,
		\begin{align}
		\Phi_-(p,e) \le \liminf_{t \uparrow  \tau}w(t,p_t,e_t) \le \limsup_{t \uparrow  \tau}w(t,p_t,e_t) \le \Phi_+(p,e),
		\end{align}
for any family $(p_t, e_t)_{0 \leq t < \tau}$ converging to $(p,e)$ as $t$ tends to $\tau$.
\end{Theorem}

\begin{Remark}
The pricing function $w$ has the remarkable property of a link between its terminal condition and its value at time $0$. Precisely, this link is shown in \eqref{infinite period w limit values} and \eqref{infinite period w limit values terminal condition} below. We explain heuristically below why this coupling emerges. It is also interesting to note that we are able to prove uniqueness of this value function in an appropriate setting; see Proposition \ref{pr uniqueness pricing function infinite model} for further details.
\end{Remark}
Let us now explain the structure of the decoupling field, $w$, in this infinite period setting.
Assume that we are given a process $E \in {\mathcal{S}}^{2,1}_{\mathrm{c}}( [0,\infty))$ such that, for any $t > 0$, $E_t$ represents the cumulative emissions up to time $t$ in the market setting introduced here. Further, assume that there is a unique price for emissions allowances in the market such that at any time $t$ with $t \neq T_k$ for any $k \geq 1$, the allowance price $Y_t$ is equal to a deterministic function, $y$, of the time $t$, the time $t$ cumulative emissions $E_t$ and the time $t$ value of the random factors in the market $P_t$ i.e. assume that there exists a deterministic function $y$ such that $Y_t = y(t, P_t, E_t)$ whenever $t \in [0, \infty) \setminus \{ T_1, T_2,....\}$. {By right continuity of $Y$, we also consider that $y(T_k,P_{T_k},E_{T_k}) = Y_{T_k +}$} and moreover
\begin{align}
\label{infinite period w limit values}
\lim_{t \uparrow T_k} y(t, P_t, E_t) = 
\begin{dcases} y(T_k, P_{T_k}, E_{T_k}), & \text{if } E_{T_k} < \Lambda_k, \\
1, & \text{if } E_{T_k} > \Lambda_k,
\end{dcases}
\end{align}
for every integer $k \geq 1$. Now, we argue that the dynamics over the period $[T_{k-1},T_k)$ are identical to the dynamics on the time period $[T_k, T_{k+1})$, except that, in the latter time period, the cap is higher by a quantity $\lambda$. The cap being $\lambda$ units larger is equivalent to the cumulative emissions being $\lambda$ units lower. Therefore, we impose that the function $y$ should satisfy
\begin{align}
\label{infinite period w limit values terminal condition}
y(T_k, p, e) = y(T_{k-1}, p, e - \lambda), 
\end{align}
for every integer $k \geq 1$ and every $(p,e) \in \R^d \times \R$. On the first period, we see that $y$ satisfies the condition \eqref{Theta k definition in introduction} and is in fact equal to $w$ for this period. We demonstrate that in fact $y$ (and thus $Y$) can be constructed from a slight modification of the value function on the first period. At the level of the pricing function, this expresses the stationarity of our setting. Let us note also that the different periods are decoupled in the sense that, for each integer $k \geq 0$, the values of the function $w$ over the time interval $[T_{k-1}, T_k)$ do not depend directly on its values over the time interval $[T_k, T_{k+1})$, thanks to the time periodicity. This is in contrast with the finite multi-period setting, but at the price of the strong coupling between initial and terminal condition seen in \eqref{Theta k definition in introduction}.


\section{The Multi - Period Model}
\label{mainresults1}


%

To prove the above results, we first need to obtain results for one-period FBSDEs of the form \eqref{general single period fbsde without p}.

\subsection{Well-posedness of one-period singular FBSDEs}

In this section, we will present results for one-period model, that will be useful in the sequel.
They are often direct extensions of the results obtained in \cite{carmona2013singularconservation}.

\vspace{2mm} A first ingredient of our study is to be able to consider terminal condition that depend on both $E$ and $P$. We now introduce the class of terminal conditions we will work with in the sequel.

\begin{Definition}\label{de cK}
Let $\cK$ be the class of functions $\phi:\R^d\times\R \rightarrow [0,1]$ such that $\phi$ is $L_\phi$-Lipschitz  in the first variable for some $L_\phi>0$ and non-decreasing in its second variable, namely
\begin{align}
|\phi(p,e)-\phi(p',e)| &\le L_\phi|p-p'| \quad\text{ for all }\quad (p,p',e) \in \R^d\times\R^d\times\R\;,
\\
\phi(p,e') &\ge \phi(p,e) \quad\text{ if }\quad e' \ge e\;,
\end{align} 
and moreover  satisfying,
\begin{align}
\sup_e \phi(p,e) = 1  \;\text{ and }\; \inf_e \phi(p,e) = 0 \quad\text{ for all }\quad p \in \R^d\;.
\end{align}
\end{Definition}

We now state an existence and uniqueness result for the one-period model with terminal condition depending on $P$ and $E$. This result is a restatement of some of the results in Chapter 2 of the thesis \cite{chotai2019thesis}. For completeness, the results have been stated in more detail in the appendix; see Proposition \ref{v existence proposition from thesis} and Theorem \ref{single period main theorem from thesis}.

\begin{Proposition} \label{pr existence uniqueness one-period} Let Assumption \ref{ass coef} hold and let $\phi$ belong to $\cK$.
Given any initial condition $(t_0,p,e) \in [0,\tau) \times \mathbb{R}^{d} \times \mathbb{R}$, there exists a unique progressively measurable 4-tuple of processes $(P^{t_0,p,e}_t, E^{t_0,p,e}_t, Y^{t_0,p,e}_t, Z^{t_0,p,e}_t)_{t_0 \leq t \leq \tau} \in \mathcal{S}^{2,d}_{\mathrm{c}}([t_0,\tau]) \times \mathcal{S}^{2,1}_{\mathrm{c}}([t_0,\tau]) \times \mathcal{S}^{2,1}_{\mathrm{c}}([t_0,\tau)) \times \mathcal{H}^{2,d}([t_0,\tau])$ satisfying the dynamics
\begin{align}  \label{general single period fbsde with p}
\begin{aligned} \mathop{}\!\mathrm{d} P^{t_0,p,e}_t &= b(P^{t_0,p,e}_t) \mathop{}\!\mathrm{d}
t + \sigma(P^{t_0,p,e}_t) \mathop{}\!\mathrm{d} W_t, & P^{t_0,p,e}_{t_0} &= p \in \mathbb{R}^{d}, \\
\mathop{}\!\mathrm{d} E^{t_0,p,e}_t &= \mu(P^{t_0,p,e}_t, Y^{t_0,p,e}_t) \mathop{}\!\mathrm{d} t, & E^{t_0,p,e}_{t_0}
&= e \in \mathbb{R}, \\ \mathop{}\!\mathrm{d} Y^{t_0,p,e}_t &= r Y^{t_0,p,e}_t
\mathop{}\!\mathrm{d} t + Z^{t_0,p,e}_{t}  \mathop{}\!\mathrm{d} {W_t}, & &
\\ \end{aligned}
\end{align}
and such that
	\begin{align}
	\label{relaxed terminal condition with p}
	\mathbb{P} \left[ \phi_{-}(P^{t_0,p,e}_\tau,E^{t_0,p,e}_\tau) \leq \lim_{t \uparrow \tau}Y^{t_0,p,e}_t \leq \phi_{+}(P^{t_0,p,e}_\tau, E^{t_0,p,e}_\tau) \right] = 1.
	\end{align}
The function defined by
\begin{align*}
[0,\tau) \times \mathbb{R}^{d} \times \mathbb{R} \ni (t_0,p,e) \rightarrow v(t_0,p,e) = Y^{t_0,p,e}_{t_0} \in \R
\end{align*}	
is continuous and satisfies
\begin{enumerate}
		\item{\label{v lipschitz constant e}For any $t \in [0,\tau)$, the function $v(t,\cdot, \cdot)$ is $1/(l_1(\tau-t))$-Lipschitz continuous with respect to $e$,}
		\item{\label{v lipschitz constant p}For any $t \in [0,\tau)$, the function $v(t,\cdot, \cdot)$ is $C$-Lipschitz continuous with respect to $p$, where $C$ is a constant depending on $L$, $\tau$ and $L_{\phi}$ only.}
		\item{\label{v condition terminal} Given $(p,e) \in \R^d \times \R$, for any family $(p_t, e_t)_{0 \leq t < \tau}$ converging to $(p,e)$ as $t \uparrow \tau$, we have
		\begin{align}\label{eq v condition terminal}
		\phi_-(p,e) \le \liminf_{t \rightarrow \tau}v(t,p_t,e_t) \le \limsup_{t \rightarrow \tau}v(t,p_t,e_t) \le \phi_+(p,e)\,.
		\end{align}
		}
\end{enumerate}

\end{Proposition}

\vspace{2mm}
\noindent The following result arises from the proof of Proposition \ref{pr existence uniqueness one-period}; see Remark \ref{approximation result remark} below.
\begin{Corollary}[Approximation result]\label{le smooth approx of v} Let Assumption \ref{ass coef} hold. Let $(\phi^n)_{n \ge 0}$ be a sequence of smooth functions belonging to $\cK$ and converging pointwise towards $\phi$ as n goes to $+\infty$. For $\epsilon>0$, consider then $v^{\epsilon,n}$ the solution to:
\begin{align}
\label{value function pde}
\partial_t u + \mu(p,u) \partial_{e} u + \cL_p u + \frac12\epsilon^2 (\partial^2_{ee}u+ \Delta_{pp}u) = ru \; \text{ and } \; u(\tau,\cdot) = \phi^n
\end{align}
where $\Delta_{pp}$ is the Laplacian with respect to $p$, and $\cL_p$ is the operator
\begin{align}
\cL_p(\varphi)(t,p,e) = \partial_p\varphi (t,p,e)b(p) + \frac{1}{2} \mathrm{Tr} \left[ a(p) \partial^{2}_{pp} \right](\varphi)(t,p,e),
\end{align}
with \textcolor{black}{$\partial_p$ denotes the Jacobian with respect to $p$},  and  $a = \sigma \sigma^{\top}$, where $\top$ is the transpose and $\partial^{2}_{p p}$ is the matrix of second derivative operators.

Then the functions $v^{\epsilon,n}$ are $C^{1,2}$ (continuously differentiable in $t$ and twice continuously differentiable in both $p$ and $e$) and $\lim_{n \rightarrow \infty} \lim_{\epsilon \rightarrow 0} v^{\epsilon,n} = v$ where the convergence is locally uniform in $[0,\tau)\times\R^d \times \R$.
\end{Corollary}

\begin{Remark}
\label{approximation result remark}
\begin{enumerate}[i)]
\item The proof of Proposition \ref{pr existence uniqueness one-period} is not given in this paper. As described in the Appendix, this result is almost identical to Theorem 2.2 and Proposition 2.10 from \cite{carmona2013singularconservation} with a similar proof. The difference between Proposition \ref{pr existence uniqueness one-period} here and the aforementioned results from \cite{carmona2013singularconservation} is that, here the terminal condition belongs to $\cK$, while in \cite{carmona2013singularconservation}, the authors considered terminal conditions $\phi: \R \rightarrow [0,1]$ which are monotone increasing, having limit $0$ at $- \infty$ and $+1$ at $+ \infty$. Moreover,
\begin{enumerate}
\item In that paper, (and also for the proof of Proposition \ref{pr existence uniqueness one-period}), the value function $v$ is constructed in the following way. First, one assumes that the coefficient functions $b$, $\sigma$ and $\mu$, and the terminal condition $\phi$ are Lipschitz smooth with bounded derivatives of all order. Then one adds to the system \eqref{general single period fbsde without p} mollifying noise of variance $\epsilon^2$ for a small $\epsilon > 0$. In this setting, results from \cite{delarue2002existence} allow one to show that the corresponding version of FBSDE \eqref{general single period fbsde without p} has a unique solution and a smooth value function satisfying a PDE of the form \eqref{value function pde}. Then, using some a priori estimates, a value function, $v$, for the FBSDE in the original setting can be obtained by taking limits, as the noise converges to $0$ and the terminal condition converges to the true, discontinuous terminal condition, of value functions for FBSDE \eqref{general single period fbsde with p} with additional mollifying noise and a smooth approximation of the true terminal condition, namely the functions $(v^{\epsilon,n})$ in Corollary \ref{le smooth approx of v}.
\item Uniqueness is obtained in \cite{carmona2013singularconservation} by a direct duality argument; see Section 2.2 therein for details. We note however that the novel estimate given in Lemma \ref{le l1 estimate one period} yields the uniqueness result too.
\end{enumerate}
\item
Corollary \ref{le smooth approx of v} presents the key approximation procedure used to construct $v$ by introducing a vanishing viscosity (and smoothing of the terminal condition). Also, note that if one heuristically considers a solution to \eqref{general single period fbsde with p} of the form $Y_t = v(t, P_t, E_t)$ with $v$ being $C^{1,2}$ then an application of It\^{o}'s formula yields the PDE \eqref{value function pde} with $\epsilon = 0$ and terminal condition replaced by $\phi$, namely
\begin{align}
\label{eq degenerate quasilinear pde}
\partial_t u + \mu(p,u) \partial_{e} u + \cL_p u  = ru \; \text{ and } \; u(\tau,\cdot) = \phi.
\end{align}
Proposition \ref{pr existence uniqueness one-period} tells us then also that $v$ can be considered as the unique ``entropy solution'' to \eqref{eq degenerate quasilinear pde} and that it can be represented using the ``random characteristics'' $(P,E,YZ)$ \cite{bressan1999hyperbolic,kruvzkov1970first}. 

\item By the results in \cite{chotai2019thesis}, Corollary \ref{le smooth approx of v} holds when the sequence $(\phi^n)_{n \geq 0}$ converges to any function $\hat{\phi}$ satisfying $\phi_{-} \leq \hat{\phi} \leq \phi_{+}$. In the thesis \cite{chotai2019thesis}, for a function $\phi \in \cK$, an explicit construction of sequences of functions $(\phi_{n, -})_{n \geq 0}$ and $(\phi_{n, +})_{n \geq 0}$, belonging to $\cK$, converging to $\phi_{-}$ and $\phi_{+}$, respectively, and with each function having the same Lipschitz constant as $\phi$, is given.
\end{enumerate}
\end{Remark}


\noindent We now collect some properties of the function $v$ defined above that will be useful in the sequel.
\begin{Proposition} \label{pr some properties 1}Let Assumption \ref{ass coef} hold. Let $v$ be the function defined in Proposition \ref{pr existence uniqueness one-period} associated to $\phi \in \cK$. Then the following hold.
\begin{enumerate}[i)]
\item {\label{it limits for v}Limits for $v$: 
\begin{align}\label{eq limit in e}
\lim_{e \rightarrow +\infty} v(t,p,e) = e^{-r(\tau-t)} \text{ and }
\lim_{e \rightarrow - \infty} v(t,p,e) = 0
\end{align}
}
\item $L^1$-integrability property: there exists a constant $C>0$ depending only on $L$ and $\tau$ such that
\begin{align}\label{eq L1 integrability}
\sup_{p \in \R^d} \int_{-\infty}^{0}v(t,p,e) \ud e \le e^{-r(\tau-t)} \left( \sup_{p\in\R^d} \int_{-\infty}^{0}\phi(p,e) \ud e + C \right)\,.
\end{align}
\item {\label{it comparaison}  Comparison property: Let $\tilde{\phi} \in \cK$ such that $\tilde{\phi}(p,e) \geq \phi(p,e)$ for every $(p,e) \in \R^d \times \R$. Then $\tilde{v}(t,p,e) \geq v(t,p,e)$ for every $(t,p,e) \in [0, \tau) \times \R^d \times \R$, where $\tilde{v}$ is defined in Proposition \ref{pr existence uniqueness one-period} with terminal condition $\tilde{\phi}$.}
\end{enumerate}
\end{Proposition}
\proof
Parts i) and iii) follow directly from Proposition \ref{v existence proposition from thesis} which is presented in the appendix.

For part ii), we use the estimate in part (\ref{v estimate in appendix}) of Proposition \ref{v existence proposition from thesis}. The following holds
\begin{align*}
  e < \Lambda \implies v(t,p,e) \leq e^{-r(\tau-t)} \left( \mathbb{E}[ \phi(P_\tau^{ t,p}, \Lambda)] + 1 \wedge \left \{ C(1+|p|^a) \left( \frac{\Lambda - e}{L(\tau-t)} \right)^{-a} \right \} \right),
\end{align*}
for $a \ge 1$ where $C$ is a constant that depends only on  $L$ and $\tau$; it may change from line to line in this proof.
Assume $\Lambda < 0$ and set  $a=2$, $e = (1+\sqrt{1+|p|^2}) \Lambda$, in the previous inequality, then
\begin{align*}
\forall \Lambda < 0, \; v(t,p,(1+\sqrt{1+|p|^2})\Lambda) \leq e^{-r(\tau-t)} \left( \esp{\phi(P_\tau^{ t,p}, \Lambda)}+ 1 \wedge \frac{C}{|\Lambda|^2}   \right)\,.
\end{align*}
Integrating on $\Lambda$, we obtain
\begin{align*}
\int_{-\infty}^{0}v(t,p,e) \ud e \leq e^{-r(\tau-t)} \left( \esp{\int_{-\infty}^{0}\phi(P_\tau^{ t,p}, e) \ud e} + C \right)\;.
\end{align*}
The proof is concluded by taking the supremum in $p$ accordingly.
\eproof

\vspace{2mm}
%

\noindent The following lemma is key to obtain uniqueness for the infinite period model. 

\begin{Lemma}\label{le l1 estimate one period} Let Assumption \ref{ass coef} hold. For $i \in \set{1,2}$, ${}^i\phi \in \cK$, we denote by ${}^iv$ the function defined in Proposition \ref{pr existence uniqueness one-period} and associated to ${}^i\phi$.
Set $\Delta v := {}^1v - {}^2v $, $\Delta \phi ={}^1\phi-{}^2\phi$. 
\\
Then, for any $(t_0,p) \in [0,\tau) \times \R^d$, it holds 
\begin{align} \label{eq l1 estimate one period}
\esp{\int |\Delta v|(t,P^{t_0,p}_t,e) \ud e} \le e^{- r ( \tau -t)} \esp{\int |\Delta \phi|(P^{t_0,p}_{\tau},e)  \ud e},
\end{align}
\textcolor{black}{for any $t \in [t_0, \tau)$, where $P^{t_0,p}$ is solution to \eqref{multi period P equation} started at $t_0$ with $P_{t_0} = p$}.
\end{Lemma}
\proof
The proof is carried out in several steps. We first make use of Corollary \ref{le smooth approx of v} and prove a form of \eqref{eq l1 estimate one period} which is valid when the functions ${}^1v$ and ${}^2v$ are replaced by approximating sequences and the integrals in \eqref{eq l1 estimate one period} are over a compact set. Then, we take limits to complete the proof. \\
1.a For $i \in \set{1,2}$,
let ${}^iv^{\epsilon,n}$ be the $C^{1,2}$ function defined in Corollary \ref{le smooth approx of v} and associated to a Lipschitz smooth approximating sequence of ${}^i\phi$. Denote ${}^iv^{n} = \lim_{\epsilon \downarrow 0} {}^iv^{\epsilon,n}$ and
let us introduce
\begin{align*}
[0,\tau]\times\R^d \ni (t,p) &\mapsto \varrho^{\epsilon,n}(t,p) := \int ({}^1v^{\epsilon,n}-{}^2v^{\epsilon,n})(t,p,e)\eta(e)\ud e
\in \R\;,
\\
[0,\tau]\times\R^d \ni (t,p) &\mapsto \varrho^{n}(t,p) := \int ({}^1v^{n}-{}^2v^{n})(t,p,e)\eta(e)\ud e
\in \R \quad \text{ and }
\\
[0,\tau]\times\R^d \ni (t,p) &\mapsto  \varrho(t,p) := \int \Delta v(t,p,e)\eta(e)\ud e \in \R,
\end{align*} 
where $\eta$ is a smooth, bounded function with compact support whose form will be chosen later. By a direct application of the dominated convergence theorem, we observe that $\lim_{\epsilon \downarrow 0} \varrho^{\epsilon,n} = \varrho^n$ and $\lim_{n \rightarrow \infty} \varrho^{n} = \varrho$. Moreover, $\varrho^{\epsilon,n}$ is a $C^{1,2}$ solution to
\begin{align}\label{eq pde difference}
\partial_t u + \cL_p u +  \frac12\epsilon^2 \partial^2_{pp}u = ru + h^{\epsilon,n}(t,p) + g^{\epsilon,n}(t,p) \; \text{ and } \; u(\tau,\cdot) = {}^1\phi^n-{}^2\phi^n
\end{align}
where
\begin{align}
[0,\tau]\times\R^d \ni (t,p) &\mapsto h^{\epsilon,n}(t,p)\! := \!\int \set{M(p,{}^1v^{\epsilon,n})-M(p,{}^2v^{\epsilon,n}}(t,p,e)\eta'(e)\ud e \in \R,
\\
[0,\tau]\times\R^d \ni (t,p) &\mapsto 
g^{\epsilon,n}(t,p) = -  \frac{1}{2} \epsilon^2 \int \set{{}^1v^{\epsilon,n}-{}^2v^{\epsilon,n}}(t,p,e)\eta''(e)\ud e
\; \in \R\,
\end{align}
{with $M(p,y) = \int_0^y\mu(p,\upsilon) \ud \upsilon$, $y \in \R$}. For later use, we remark that $g^{\epsilon,n} \rightarrow 0$ as $\epsilon \rightarrow 0$ and that
\begin{align*}
\lim_{n \rightarrow \infty} \lim_{\epsilon \downarrow 0} h^{\epsilon,n}(t,p) = h^\eta(t,p):=\int \set{M(\cdot,{}^1v)-M(\cdot,{}^2v}(t,p,e)\eta'(e)\ud e.
\end{align*}
Let $(P^\epsilon_t)_{t \in [t_0,T]}$ be the solution to
\begin{align*}
P^\epsilon_t = p + \int_{t_0}^t b(P^\epsilon_s) \ud s +  \int_{t_0}^t \sigma(P^\epsilon_s) \ud W_s  + \epsilon (W'_s-W'_{t_0}),
\end{align*},
where $W'$ is a Brownian motion independent from $W$,
and observe that, by classical arguments, $\lim_{\epsilon \downarrow 0}\esp{\sup_{t \in [t_0, \tau]}|P^{\epsilon}_t - P^{t_0,p}_t|} = 0$. Since $(\varrho^{\epsilon,n})_{\epsilon>0}$ are uniformly Lipschitz, we straightforwardly deduce that
\begin{align*}
\lim_{\epsilon \downarrow 0} \esp{|\varrho^{\epsilon,n}(t,P^{\epsilon}_t) - \varrho^{n}(t,P^{t_0,p}_t) |^2} = 0\;.
\end{align*}
Now,  we apply It\^{o}'s Formula to $(e^{-r t}\varrho^{\epsilon,n}(t,P^\epsilon_t))_{0\le t < \tau}$. Using the PDE \eqref{eq pde difference}, we get
\begin{align*}
\esp{e^{-r t}\varrho^{\epsilon,n}(t,P^\epsilon_t)} = 
e^{-r t_0}\varrho^{\epsilon,n}(t_0,p) + \esp{\int_{t_0}^t 
		e^{-r s}(h^{\epsilon,n}+ g^{\epsilon,n})(s,P^\epsilon_s) \ud s} \;.
\end{align*}
Taking the limit in $\epsilon$ first and then $n$ (recall uniform linear growth in $p$), we obtain
\begin{align*}
\esp{ e^{-r t}\varrho(t,P^{t_0,p}_t) }=
 e^{-r t_0}\varrho(t_0,p) + \esp{ \int_{t_0}^t  e^{-r s} h^\eta(s,P^{t_0,p}_s)  \ud s}
\end{align*}
To conclude this step, we observe that using \eqref{eq v condition terminal}, it follows from the dominated convergence theorem
\begin{align*}
\lim_{t \rightarrow \tau} \int v(t,p,e)\eta(e) \ud e = \int \phi(p,e)\eta(e) \ud e.
\end{align*}
Combining the above observation with the dominated convergence theorem again, we get
\begin{align*}
\lim_{t \rightarrow \tau}\esp{\varrho(t,P_t)}  = \esp{\int \Delta \phi(P_{\tau},e)\eta(e) \ud e}\;.
\end{align*}
We thus have
\begin{align}\label{eq lemma L1 temp}
\esp{\int \Delta v(t,P_t,e)\eta(e) \ud e} = \esp{e^{- r ( \tau -t)} \int \Delta \phi(P_{\tau},e)\eta(e) \ud e
+ \int_t^{\tau} e^{- r ( s -t)}h^\eta(s,P_s)\ud s}
\end{align}
1.b Note that the previous reasoning can be applied to $\hat{\phi}={}^1\phi \wedge {}^2\phi \in \cK$ and $\check{\phi}={}^1\phi \vee {}^2\phi \in \cK$ and the terminal condition $\widetilde{\Delta \phi} :=  \check{\phi}-\hat{\phi} := |{}^1\phi - {}^2\phi|$. Denoting $\hat{v}$ and $\check{v}$ the associated functions, we have from the comparison result given in Proposition \ref{pr some properties 1} (\ref{it comparaison}), that $\hat{v} \le {}^1v \wedge {}^2v $ and $\check{v} \geq {}^1v \vee {}^2v$ and then $\check{v}-\hat{v} \ge |\Delta v|$. Combined with \eqref{eq lemma L1 temp}, this leads to
\begin{align}
\label{eq lemma L1 temp 1.b}
\esp{\int |\Delta v(t,P_t,e)|\eta(e) \ud e} \le  \mathbb{E} \bigg[ e^{- r ( \tau -t)}\int |\Delta \phi|(P_{\tau}, e)\eta(e) \ud e
	+ \int_t^{\tau} |h^\eta|(s,P_s)\ud s \bigg] \,.
\end{align}
\\
2. We recall that for $\phi \in \cK$, for all $(t,p) \in [0, \tau]\times\R^d$, we have
\begin{align}\label{eq limit in e in lemma}
\lim_{e \rightarrow +\infty} v(t,p,e) = e^{-r(\tau -t)} \text{ and }
\lim_{e \rightarrow - \infty} v(t,p,e) = 0;
\end{align}
see Proposition \ref{pr some properties 1} \eqref{it limits for v}.\\
For this step, we let $\eta$  be a smooth approximation of $1_{[-R,R]}$ for $R>0$ with first and second derivatives bounded (uniformly in $R$), equal to $1$ on $[-R,R]$  and with support in ${[-(R+1),(R+1)]}$(in other words a smooth truncation function). \\
Observe that, for all $(t,p) \in [0,\tau] \times \R^d$, 
\begin{multline*}
\lim_{R \rightarrow +\infty}\1_{[R,R+1]}(e)|M(p,{}^1{v})(t,p,e)-M(p,{}^2{v})(t,p,e)| 
=
\\
|\lim_{e \rightarrow +\infty} M(p,{}^1{v})(t,p,e)- \lim_{e \rightarrow +\infty} M(p,{}^2{v})(t,p,e)| =0,
\end{multline*}
where we used \eqref{eq limit in e in lemma}. Similarly,
\begin{align*}
\lim_{R \rightarrow +\infty}\1_{[-R-1,-R]}(e)|M(p,{}^1{v})(t,p,e)-M(p,{}^2{v})(t,p,e)| = 0\,.
\end{align*}
Observing that $|M(p,{}^1{v})(t,p,e)| \le L(2 + |p|)$,
we deduce from the dominated convergence theorem and the previous observations that
\begin{align}\label{eq limit heta}
\lim_{R \rightarrow +\infty}\esp{\int_{t_0}^{\tau} |h^\eta|(s,P^{t_0,p}_s)\ud s} = 0\;.
\end{align}
Moreover, the monotone convergence theorem implies that
\begin{align*}
&\lim_{R \rightarrow +\infty} \esp{\int |\Delta v(t,P^{t_0,p}_t,e)|\eta(e) \ud e}  = \esp{\int |\Delta v(t,P^{t_0,p}_t,e)| \ud e} ,
\\
&\text{and}  \lim_{R \rightarrow +\infty} \esp{\int |\Delta \phi|(P^{t_0,p}_{\tau},e) \eta(e) \ud e} = \esp{\int |\Delta \phi|(P^{t_0,p}_{\tau},e) \ud e}\;.
\end{align*}
The proof is concluded by combining the above equalities with \eqref{eq limit heta} and \eqref{eq lemma L1 temp 1.b}.
\eproof

\begin{Lemma}\label{le control lipschitz}
Consider $\phi \in \cK$ and assume that Assumptions \ref{ass coef}, \ref{ass cap infinite model} and \ref{ass infinite model coef} hold. Then, the corresponding function $v$, defined in Proposition \ref{pr existence uniqueness one-period}, satisfies, for all $(t,p,p',e)\in [0,\tau)\times\R^d \times \R^d \times \R$,
\label{le control lipschitz} 
\begin{align}
\label{v multi period with assumptions bound}
|v(t,p,e)-v(t,p',e)| \le \left(e^{-\beta(\tau -t)}L_\phi +   \frac{L}{l_1}\right) |p-p'|\,,
\end{align}
where $L_\phi$ is such that
$
|\phi(p,e)-\phi(p',e)| \le L_\phi |p-p'|\;.
$
\end{Lemma}
\proof 
We split the proof into two steps. In the first step, we use the PDE \eqref{value function pde} to prove \eqref{v multi period with assumptions bound} in a setting in which all coefficient functions are smooth with bounded derivatives of all orders, the terminal condition $\phi$ is smooth and the FBSDE in Proposition \ref{pr existence uniqueness one-period} has additional mollifying noise. Then, we use a standard mollification argument and Corollary \ref{le smooth approx of v} to conclude.

\noindent 1. Using a mollification argument, the same as that used in Section 2 of \cite{carmona2013singularconservation}, we start by assuming that the coefficient functions $b$, $\sigma$ and $\mu$ are smooth with bounded derivatives of all orders. We also assume that the terminal condition $\phi$ is Lipschitz smooth. In this setting, following Section 2 of \cite{carmona2013singularconservation}, given $\epsilon > 0$, we can associate with the FBSDE a function $v^{\epsilon}$ which is a smooth solution to the PDE in \eqref{value function pde} with terminal condition $v^{\epsilon}(\tau, \cdot, \cdot) = \phi(\cdot, \cdot)$. The function $v^{\epsilon}$ has bounded and continuous derivatives of any order on $[0, \tau] \times \R^d \times \R$.
\\
\\
Without loss of generality, we can consider only the first component of the gradient of $v^\epsilon$: $w^\epsilon := {\partial_{p_1}} v^\epsilon$. The proof will be similar for all other components. Using the PDE \eqref{value function pde}, we have that
\begin{multline}
\label{eq gradient v}
\partial_t w^\epsilon + \tilde{b}(p)\cdot\partial_p w^\epsilon + \frac12 Tr[\set{a(p) + \epsilon^2 I_d} \partial^2_{pp}w^\epsilon] + \frac{1}{2} \epsilon^2 \partial^{2}_{ee} w^\epsilon +  b'_1(p) w^\epsilon  \\
+ \mu(p,v^\epsilon) \partial_e w^\epsilon + [\partial_{p_1} \mu(p,v^\epsilon) + \partial_y \mu(p,v^\epsilon) w^\epsilon] \partial_e v^\epsilon = r w^\epsilon 
\end{multline}
with, $\tilde{b}(p)$ is a vector the following components: for $1 \le i \le d$,
\begin{align*}
\begin{dcases}
\tilde{b}_i(p) := b_i(p) + \sigma'_{1\cdot}(p)^\top \sigma_{i\cdot}(p) , & {\text{if }i = 1}, \\
\tilde{b}_i(p) := b_i(p) + \frac12 \sigma'_{1\cdot}(p)^\top \sigma_{i\cdot}(p), &{\text{otherwise,} } 
\end{dcases}
\end{align*}
and, for $1 \leq i \leq d$, we are denoting $b^{\prime}_{i}(p) = \partial_{p_i} b_i(p)$ and $\sigma^{\prime}_{i \cdot}(p) = \partial_{p_i} \sigma_{i \cdot}(p)$, where $\sigma_{i \cdot}(p)$ denotes row $i$ of the matrix $\sigma(p)$. Now set, for \sloppy$t \in [t_0, \tau]$, $(U^\epsilon_t,V^\epsilon_t)  = (w^\epsilon(t,\tilde{P}^\epsilon_t,\tilde{E}^\epsilon_t),\partial_p w^\epsilon(t,\tilde{P}^\epsilon_t,\tilde{E}^\epsilon_t))$ with $(\tilde{P}^\epsilon,\tilde{E}^\epsilon)$ strong solution to
\begin{align}
\ud \tilde{P}^\epsilon_t &= \tilde{b}(\tilde{P}^\epsilon_t) \ud t + \sigma(\tilde{P}^\epsilon_t) \ud W_t + \epsilon \ud W'_t\;,\;  \tilde{P}^\epsilon_{t_0} = p\,,
\\
\ud \tilde{E}^\epsilon_t &= \mu(\tilde{P}^\epsilon_t, v^\epsilon(t,\tilde{P}^\epsilon_t,\tilde{E}^\epsilon_t)) \ud t + \epsilon \ud B_t\;,\;  \tilde{E}^\epsilon_{t_0} = e\,,
\end{align}
where $(W',B)$ is two dimensional Brownian Motion independent from $W$.
Applying It\^{o}'s formula, we compute
\begin{multline*}
\begin{aligned}
\ud U^\epsilon_t &= \set{r - b'_1(\tilde{P}^\epsilon_t)}U^\epsilon_t \ud t - \partial_{p_1}\mu\left(\tilde{P}^\epsilon_t, v^\epsilon(t,\tilde{P}^\epsilon_t,\tilde{E}^\epsilon_t)\right)\partial_e v^\epsilon(t,\tilde{P}^\epsilon_t,\tilde{E}^\epsilon_t) \ud t
\\
& \quad -\partial_y \mu\left(\tilde{P}^\epsilon_t, v^\epsilon(t,\tilde{P}^\epsilon_t,\tilde{E}^\epsilon_t)\right)\partial_e v^\epsilon(t,\tilde{P}^\epsilon_t,\tilde{E}^\epsilon_t) U^\epsilon_t\ud t
\\
& \quad + V^\epsilon_t \sigma(\tilde{P}^\epsilon_t) \ud W_t + \epsilon V^\epsilon_t \ud W'_t + \epsilon \partial_e w(t,\tilde{P}^\epsilon_t,\tilde{E}^\epsilon_t) \ud B_t.
\end{aligned}
\end{multline*}
We now introduce the weights, 
\sloppy
\begin{align*}
\cI_t =  \exp \Big( \int_{t_0}^t\big(b'_1(\tilde{P}^\epsilon_s)-r \big) \ud s \Big)\,,
\hspace{0.2em} 
\cJ_t = \exp \Big( \int_{t_0}^t\partial_y \mu \big(\tilde{P}^\epsilon_t, v^\epsilon(t,\tilde{P}^\epsilon_t,\tilde{E}^\epsilon_t)\big) \partial_e v(s,\tilde{P}^\epsilon_t,\tilde{E}^\epsilon_t) \ud s \Big) \,,
\end{align*}
$\text{ and }\cE_t = \cI_t \cJ_t$ for $t \in [t_0, \tau]$.
Setting $(\bar{U}^\epsilon_t, \bar{V}^\epsilon_t) = \cE_t (U^\epsilon_t,V^\epsilon_t)$ for $t \in [t_0, \tau]$, we compute
\begin{align*}
\ud \bar{U}^\epsilon_t &= - \partial_{p_1}\mu\left(\tilde{P}^\epsilon_t, v^\epsilon(t,\tilde{P}^\epsilon_t,\tilde{E}^\epsilon_t)\right)\partial_e v^\epsilon(t,\tilde{P}^\epsilon_t,\tilde{E}^\epsilon_t) \cE_t \ud t
\\
&+ \bar{V}^\epsilon_t \sigma(\tilde{P}^\epsilon_t) \ud W_t + \epsilon \bar{V}^\epsilon_t \ud W'_t + \epsilon \partial_e w(t,\tilde{P}^\epsilon_t,\tilde{E}^\epsilon_t) \cE_t \ud B_t.
\end{align*}
This leads to
\begin{align*}
\esp{\bar{U}^\epsilon_{t_0}} & =\esp{\bar{U}^\epsilon_{\tau}} + \esp{\int_{t_0}^{\tau}\partial_{p_1}\mu\left(\tilde{P}^\epsilon_t, v^\epsilon(t,\tilde{P}^\epsilon_t,\tilde{E}^\epsilon_t)\right)\partial_e v^\epsilon(t,\tilde{P}^\epsilon_t,\tilde{E}^\epsilon_t) \cE_t \ud t},
\\
&=\esp{\bar{U}^\epsilon_{\tau}} + \esp{\int_{t_0}^{\tau} \cI_t 
	\frac{\partial_{p_1}\mu\left(\tilde{P}^\epsilon_t, v^\epsilon(t,\tilde{P}^\epsilon_t,\tilde{E}^\epsilon_t)\right) }
	{\partial_{y}\mu\left(\tilde{P}^\epsilon_t, v^\epsilon(t,\tilde{P}^\epsilon_t,\tilde{E}^\epsilon_t)\right) }
	\dot{\cJ}_t \ud t}.
\end{align*}
By Assumptions \ref{ass coef} and \ref{ass infinite model coef}, we have  $|\frac{\partial_{p_1}\mu}{\partial_{y}\mu}| \le \frac{L}{\ell_1}$ and $\cI_t \le 1$. Therefore
\begin{align*}
|\partial_{p_1}v^\epsilon(t_0,p,e)| \le \esp{\cE_{\tau}}|\partial_{p_1}\phi|_\infty  + \frac{L}{\ell_1}(1-\esp{\cJ_{\tau}}).
\end{align*}
Using the fact that $r - |b'_1|_\infty \ge \beta$ and $\cJ_{\cdot} \le 1$, we get
\begin{align*}
|\partial_{p_1}v^\epsilon|_\infty \le e^{-\beta(\tau -t_0)}|\partial_{p_1}\phi|_\infty +   \frac{L}{\ell_1}\;,
\end{align*}
which concludes the proof for this step.
\\
\noindent 2. Using the mollification argument, we can approximate the true coefficient functions $b$, $\sigma$ and $\mu$ by smooth functions with bounded derivatives of all orders and take limits to show that \eqref{v multi period with assumptions bound} will hold for $v^{\epsilon}$ without any additional smoothness or boundedness conditions on the coefficient functions.

\noindent Finally, for a general terminal condition $\phi \in \cK$, we can approximate $\phi$ by a sequence of Lipschitz smooth functions in $\cK$ and apply Corollary \ref{le smooth approx of v} to complete the proof.
\eproof


\begin{Lemma} \label{le other control lipschitz}
Consider $\phi \in \cK$ and assume that Assumptions \ref{ass coef}, \ref{ass cap infinite model} and \ref{ass unif ellip}(i) hold. Then, the corresponding function $v$, defined in Proposition \ref{pr existence uniqueness one-period}, satisfies, for all $(t,p,p',e)\in [0,\tau)\times\R^d \times \R^d \times \R$,
\begin{align}
\label{v multi period with assumptions bound}
|v(t,p,e)-v(t,p',e)| \le C_{\beta}\left( \frac{\|\phi\|_\infty}{\sqrt{\tau-t}} +   \frac{1}{l_1}\right) |p-p'| .
\end{align}
\end{Lemma}

\proof
We proceed as in the proof of Lemma \ref{le control lipschitz}: We consider first the noisy version of the system (for a small parameter $\epsilon>0$) with smooth terminal condition and coefficient functions, namely
\begin{align}
P^{t_0,p}_t &= p + \int_{t_0}^t b(P^{t_0,p}_s)\ud s + \sum_{\ell=1}^d \int_{t_0}^t \sigma^{\cdot\ell}(P^{t_0,p}_s)\ud W^\ell_s 
\\
E^{t_0,p,e,\epsilon}_t &= e +\int_{t_0}^t \mu(P^{t_0,p}_s,E^{t_0,p,e,\epsilon}_s)\ud s + \epsilon B_s
\\
Y_t^{t_0,p,e,\epsilon} &= \phi(P^{t_0,p}_\tau,E^{t_0,p,e,\epsilon}_\tau) - \int_t^\tau Z^{t_0,p,e,\epsilon}_s \ud W_s \label{eq de Yeps}
\end{align}
where $B$ is a Brownian Motion independent from $W$ and, in this smooth setting, we have that for $t_0 \le t \le \tau$, $Y^\epsilon_t = v^\epsilon(t,P^{t_0,p}_t,E^{t_0,p,e,\epsilon}_t)$, $Z_t^\epsilon= \partial_p v^\epsilon(t,P^{t_0,p}_t,E^{t_0,p,e,\epsilon}_t) \sigma(P^{t_0,p}_t)$, where $v^\epsilon$ is a classical solution to
\begin{align}\label{eq pde for veps}
\partial_t v^\epsilon + \partial_p v^\epsilon b(p) + \frac12Tr[a(p)\partial^2_{pp}v^\epsilon] + \frac{\epsilon^2}2 \partial^2_{ee}v^\epsilon + \mu(p,v^\epsilon)\partial_e v^\epsilon= r v^\epsilon \text{ and } v^\epsilon(\tau,\cdot)= \phi(\cdot)\;.
\end{align}
Note that we do not need to add noise on the $P$-component as $\sigma$ is assumed to be uniformly elliptic here, see Assumption \ref{ass unif ellip}(i).
In particular,  we deduce from \eqref{eq pde for veps}, that for $1 \le i \le d$, $\partial_{p_i}v^\epsilon$ satisfies the following equation:
\begin{align}
r u &= \partial_t u + \partial_p u b(p) + \frac12Tr[a(p)\partial^2_{pp}u] + \frac{\epsilon^2}2 \partial^2_{ee}u + \mu(p,v^\epsilon)\partial_e u \noindent
\\
&+ \partial_p v^\epsilon \partial_{p_i}b(p)+\frac12 Tr[\partial_{p_i}a(p)\partial^2_{pp}v^\epsilon]
+ \partial_{p_i} \mu(p,v^\epsilon)\partial_e u +\partial_y \mu(p,v^\epsilon)\partial_e  u\,.
\label{eq pde for d_pv}
\end{align}
We also introduce the tangent process associated to $P^{t_0,p}$ valued in the set of $d\times d$ matrices:
\begin{align}\label{de tangent P}
\partial_p P^{t_0,p}_t = I_d + \int_0^t\partial_pb(P^{t_0,p}_s) \partial_p P^{t_0,p}_s \ud s + \sum_{\ell=1}^d \int_0^t\partial_p \sigma^{\cdot\ell}(P^{t_0,p}_s)\partial_pP^{t_0,p}_s \ud W^\ell_s\;,
\end{align}
where for $1 \le k \le d$, we have $(\partial_p P^{t_0,p}_t )^{.k} = \partial_{p_k} P^{t_0,p}$ and $I_d$ is the identity matrix.
The following estimate is well known, see e.g. \cite{ma2002representation}: for $\kappa \ge 1$,
\begin{align}\label{eq bound partial P}
\esp{\sup_{t\in[t_0,\tau]}|\partial_p P^{t_0,p}_t|^\kappa} \le C_{\kappa}\;.
\end{align}
$C_{\kappa}$ depends only on $T$ and $L$ the Lipschitz constant of $b$, $\sigma$ and the extra parameter $\kappa$.
\\
Below, we drop the dependence in $(t_0,p,e)$ for the reader's convenience. The remainder of the proof is split into two steps: we first obtain a Bismut-Elworthy-Li type formula for the gradient of $v^{\epsilon}$ with respect to $p$ and then use it to obtain the desired upper bound.
\\
1.a Applying Ito's formula to $U^\epsilon = \partial_{p} v^\epsilon(t,P_t,E^\epsilon_t)$, one gets
\begin{align*}
\ud U^\epsilon_t & = - U^\epsilon_t \Big( \partial_p b(P_t)
+(r+\partial_y \mu(P_t,Y^\epsilon_t)\partial_e v^\epsilon(t,P_t,E^\epsilon_t)  )I_d
\Big) \ud t - \frac12 A^\epsilon_t \ud t
\\
&+ \partial_p \mu(P_t,Y^\epsilon_t)\partial_e v^\epsilon(t,P_t,E^\epsilon_t) \ud t
+ \epsilon \partial^2_{pe} v^\epsilon(t,P_t,E^\epsilon_t) \ud B_t
+ \ud M^{\epsilon}_t
\end{align*}
where 
\begin{align*}
A^\epsilon_t &= ({Tr} \left[\partial_{p_\ell}a(P_t)V^\epsilon_t \right])_{1 \le \ell \le d} ,\\ 
\ud M^{\epsilon}_t &= ((V_t^\epsilon)^{\ell \cdot}\sigma(P_t)\ud W_t)_{1 \le \ell \le d},
\end{align*} 
and 
\begin{align*}
V^\epsilon_t = \partial^{2}_{pp} v^\epsilon(t,P_t,E^\epsilon_t).
\end{align*}
Introducing,
\begin{align}
\cE_t :=  \exp \Big( \int_{t_0}^t\partial_y \mu \big(P_t, v^\epsilon(t,P_t,{E}^\epsilon_t)\big) \partial_e v^\epsilon(s,{P}_t,{E}^\epsilon_t) \ud s \Big)\;,
\end{align}
we then apply Ito's formula to $(\bar{U}^\epsilon_t := e^{-r (t-t_0)} \cE_t U^\epsilon_t \partial_p P_t)_{t_0 \le t \le \tau}$, to get
\begin{align}
\ud \bar{U}^\epsilon_t =  e^{-r(t-t_0)}\partial_p \mu(P_t,Y^\epsilon_t)\partial_p P_t \,\cE_t \partial_e v^\epsilon(t,P_t,E^\epsilon_t) \ud t
+ \ud \cM^\epsilon_t\,,
\end{align}
where $\cM^\epsilon$ is a square integrable martingale. In particular, we observe that, for $t_0 \le t \le \tau$,
\begin{align} \label{eq rep step 1}
\partial_p v^\epsilon(0,p,e) &= \esp{e^{-r(t-t_0)} \cE_t U^\epsilon_t \partial_p P_t
+ \Theta_t}
\end{align}
\text{ with } 
\begin{align}\label{eq de Theta}
\Theta_t &= \int_{t_0}^t e^{-rs}\partial_p \mu(P_s,Y^\epsilon_s)\partial_p P_s \,\cE_s\partial_e v^\epsilon(s,P_s,E^\epsilon_s) \ud s \;.
\end{align}
Let us introduce
\begin{align}
H^{t_0}_\tau := \frac1{\tau-t_0} \left(\int_{t_0}^\tau\left(e^{-r(t-t_0)} \cE_t  \sigma(P_t)^{-1}\partial_p P_t \right)^\top \ud W_t \right)^{\!\top},
\end{align}
which is a random $d$-dimensional row vector.
Integrating \eqref{eq rep step 1}, from $t_0$ to $\tau$ and using Ito Isometry,
we get
\begin{align*}
\partial_p v^\epsilon(t_0,p,e) &= \esp{  \int_{t_0}^\tau U^\epsilon_t \sigma(P_t)\ud W_tH^{t_0}_\tau + \frac1{\tau-t_0} \int_{t_0}^\tau\Theta_t \ud t}.
\end{align*}
Observing that $Z^\epsilon = U^\epsilon \sigma(P)$ and using \eqref{eq de Yeps}, we deduce from the previous equality
\begin{align}\label{eq rep partial p v}
\partial_p v^\epsilon(t_0,p,e) = \esp{\phi(P_\tau,E^\epsilon_\tau)H^{t_0}_\tau +  \frac1{\tau-t_0} \int_{t_0}^\tau \Theta_t \ud t}.
\end{align}
1.b We now use \eqref{eq rep partial p v} to give an upper bound on $\|\partial_p v^\epsilon \|_\infty$. We first observe that, 
\begin{align}\label{eq bound discount}
0 \le e^{-r(t-t_0)}\cE_t \le 1\,\; \text{ for all $t_0 \le t \le \tau$},
\end{align} 
as $\partial_y \mu(\cdot) \le 0$, recall Assumption \ref{ass coef}, and $\partial_e v^\epsilon(\cdot) \ge 0$.\\
Combing \eqref{eq bound partial P}, with Assumption \ref{ass unif ellip}(i) and \eqref{eq bound discount}, we obtain
classically
\begin{align}\label{eq classic bound H}
\esp{|H^{t_0}_\tau|^\kappa}^{\frac1\kappa} \le \frac{C_\kappa}{\sqrt{\tau-t_0}}\;, \text{ for all } \kappa \ge1\;.
\end{align}
We also compute, using \eqref{eq bound partial P} and Assumption \ref{ass coef},
\begin{align}
\esp{|\Theta_t|} \le \frac{C}{l_1} \int_{t_0}^{\tau} \cE_s \partial_y \mu \big(P_t, v^\epsilon(t,P_t,{E}^\epsilon_t)\big) \partial_e v^\epsilon(s,P_s,E^\epsilon_s) \ud s\;,
\end{align}
Now, observing that $\cE_s \partial_y \mu \big(P_t, v^\epsilon(t,P_t,{E}^\epsilon_t)\big) \partial_e v^\epsilon(s,P_s,E^\epsilon_s)= \dot{\cE}_s$, we obtain
\begin{align} \label{eq bound Theta}
\esp{|\Theta_t|} \le \frac{C}{l_1}\;.
\end{align}
Combining \eqref{eq bound Theta} and \eqref{eq classic bound H} with \eqref{eq rep partial p v}, we conclude for this step
\begin{align}
\|\partial_p v^\epsilon(t_0,\cdot)\|_{\infty} \le C\Big(\frac{1}{l_1} + \frac{\|\phi\|_\infty}{\sqrt{\tau-t_0}}\Big)\;.
\end{align}
The proof is concluded by invoking the local uniform convergence of $v^\epsilon$ to $v$ when removing the noise and smoothness of the coefficient functions.
\eproof

\begin{Remark} Let us mention that \eqref{eq rep partial p v} is a Bismut-Elworthy-Li formula for this partially degenerate setting and it differs from the representations obtained in Proposition 5.1 in \cite{carmona2013singularconservation}. It is well expected since we impose non-degeneracy of the $P$-component.
Moreover, the control obtained in Lemma \ref{le other control lipschitz} indicates that existence could be obtained with less regularity than the Lipchitz assumption on $\phi$ in the $p$-variable imposed in Proposition \ref{pr existence uniqueness one-period}. 
\end{Remark}

\subsection{Parametrised one-period FBSDE}

In the sequel, we  need to consider terminal conditions parametrised by an extra variable, representing the initial value of the emissions process.
%
\begin{Definition}\label{de hat cK}
Let $\widehat{\cK}$ be the class of bounded measurable functions
$$\R^d\times \R \times \R \ni (p,e,\mathfrak{e})\rightarrow \Phi(p,e,\mathfrak{e}) \in \R$$ 
such that,
\begin{itemize}
	\item for each $\mathfrak{e} \in \R$, $\Phi(\cdot, \cdot, \mathfrak{e}) \in \cK$ with a Lipschitz constant $L_{\Phi}$ that is independent of $\mathfrak{e}$ i.e. there is $L_{\Phi} > 0$ such that
		\begin{align}
		|\Phi(p_1, e, \mathfrak{e}) - \Phi(p_2, e, \mathfrak{e})| \leq L_{\Phi} |p_1 - p_2|,
		\end{align}
		for every $p_1, p_2 \in \R^d$ and every $e, \mathfrak{e} \in \R$.
	\item for each $e \in \R$ the function
	$$
	\R^d \times \R \ni (p, \mathfrak{e}) \mapsto \Phi(p,e + \mathfrak{e},\mathfrak{e}) \in \R
	$$
	is in $\cK$.
\end{itemize} 
\end{Definition}
\noindent Note that there is a natural injection from $\cK$ into $\widehat{\cK}$: Indeed, it is clear that for every function $\phi \in \cK$, the function $\widehat{\phi} \in \widehat{\cK}$ given by
\begin{align}
\R^d \times \R \times \R \ni (p,e, \mathfrak{e}) \mapsto \widehat{\phi}(p,e, \mathfrak{e}) = \phi(p,e) \in \R\,,
\end{align}
belongs to $\widehat{\cK}$. 

\begin{Remark}
\label{re typical term condition}
In the multi period model, the cap on emissions in a given period may depend on the level of emissions made up until the start of that period. In a typical setup, we may have a regulator releasing a pre-determined quantity of allowances at the start of every period. The cap on emissions in a period is then determined by the total number of allowances available in the market at the time of compliance, taking into account all emissions made and allowances already surrendered in previous periods. The typical form of a terminal condition $\Phi$ in $\widehat{\cK}$ is
\begin{align}
\label{phi typical K hat}
\Phi(p, e, \mathfrak{e}) = \begin{dcases}
h(p,e), & \text{if } e <  \Lambda( \mathfrak{e}), \\
1, & \text{otherwise.}
\end{dcases}
\end{align}
Here, $\mathfrak{e}$ can be considered to be a parameter representing the total recorded emissions at the start of the compliance period. The function $\Lambda$ maps this quantity to the total number of allowances available in the market at the start of the period. This function accounts for the regulator's releases of allowances, see e.g. Example \ref{simple cap example}. So, $\Lambda( \mathfrak{e} )$ represents the number of allowances available to use for compliance and is therefore the cap on emissions. In \eqref{phi typical K hat}, $p$ and $e$ will represent, respectively, the value of the noise process and of the cumulative emissions process, both at the end of the compliance period. The penalty here is, as stated before, set equal to $1$, and it is incurred when total emissions exceed the cap. The function $h$ represents the value that the allowance price will take at the start of the following compliance period if over-emission has not occurred. Typically, $h$ is determined by the dynamics of the following period. In this paper, we want to consider terminal conditions of the form \eqref{phi typical K hat} with $\Lambda \in \Theta$ and $h$ satisfying appropriate conditions inherited from the properties of the value function for the FBSDE for a single period model and allow us to prove, in a recursive manner, that the multi period pricing problem is well posed. This is what motivates the definition of $\widehat{\cK}$.
\end{Remark}

The following result is the version of Proposition \ref{pr existence uniqueness one-period} for terminal conditions depending on a parameter, $\mathfrak{e}$. It follows directly from the proof of Proposition \ref{pr existence uniqueness one-period}.
\begin{Corollary} 
\label{co one period general} 
Let Assumption \ref{ass coef} hold. For $\Phi \in \hat{\cK}$,
there exists a unique function $v^\Phi$ defined by
\begin{align*}
[0,\tau) \times \mathbb{R}^{d} \times \mathbb{R}\times \mathbb{R}  \ni (t_0,p,e,\mathfrak{e}) \rightarrow v^\Phi(t_0,p,e,\mathfrak{e}) = Y^{t_0,p,e,\mathfrak{e}}_{t_0} \in \R,
\end{align*}	
where $(P^{t_0,p},E^{t_0,p,e,\mathfrak{e}},Y^{t_0,p,e,\mathfrak{e}})$ is a solution to \eqref{general single period fbsde with p}-\eqref{relaxed terminal condition with p} with terminal condition $\Phi(\cdot, \cdot \mathfrak{e})$.
\\
It is continuous and satisfies
\begin{enumerate}
		\item{\label{v lipschitz constant e in corollary}For any $t \in [0,\tau)$, the function $v^\Phi(t,\cdot, \cdot, \cdot)$ is $1/(l_1(\tau-t))$-Lipschitz continuous with respect to $e$.}
		\item{\label{v lipschitz constant p in corollary}For any $t \in [0,\tau)$, the function $v^\Phi(t,\cdot, \cdot,\cdot)$ is $C$-Lipschitz continuous with respect to $p$, where $C > 0$ depends on $L$, $L_{\Phi}$ and $\tau$ only.
		\item{\label{v condition terminal in corollary} Given $(p,e) \in \R^d \times \R$ and $\mathfrak{e} \in \R$, for any family $(p_t, e_t)_{0 \leq t < \tau}$ converging to $(p,e)$ as $t \uparrow \tau$, we have
		\begin{align}
		\Phi_-(p,e,\mathfrak{e}) \le \liminf_{t \rightarrow \tau}v^\Phi(t,p_t,e_t,\mathfrak{e}) \le \limsup_{t \rightarrow \tau}v^\Phi(t,p_t,e_t,\mathfrak{e}) \le \Phi_+(p,e,\mathfrak{e})
		\end{align}
		uniformly for $p$ in compact subset of $\R^d$.
		}
		\item{\sloppy For any  parameters $(t_0,p,e,\mathfrak{e}) \in [0,\tau)\times \mathbb{R}^{d} \times \mathbb{R}\times \mathbb{R}$, 
		$v^\Phi(t,P^{t_0,p}_t,E^{t_0,p,e,\mathfrak{e}}_t,\mathfrak{e}) = Y^{t_0,p,e,\mathfrak{e}}_t$ for $t_0\le t< \tau$ where $(P^{t_0,p},E^{t_0,p,e,\mathfrak{e}},Y^{t_0,p,e,\mathfrak{e}})$.}
}
\end{enumerate}
\end{Corollary}


\noindent The following lemma is key to show the existence of a solution to the multi-period model.
\begin{Lemma}\label{le monotonicity for induction} 
Let $\Phi \in \hat{\cK}$ and $v^\Phi$ be given by Corollary \ref{co one period general}. Then, for every $t \in [0,\tau), p \in \R^d$, the function
			\begin{align*}
		e \mapsto v^\Phi(t,p,e,{e}),
		\end{align*} 
is monotone increasing and satisfies
	\begin{align}
	\label{v with parameter limit in lemma}
\lim_{e \rightarrow - \infty} v^\Phi(t,p,e,{e}) = 0.
\end{align}	
\end{Lemma}
\proof
	We split the proof into two steps. We first use a change of variables to prove the monotonicity property and then use bounds for value functions for singular FBSDEs to prove the limit property \eqref{v with parameter limit in lemma}. \\
1. For a given $t_0 \in [0,\tau)$, $p \in \R^d$ and $e_0 \in \R$, we have $v^{\Phi}(t_0, p, e_0, e_0) = Y_{t_0}$, where $Y$ is part of the solution $(P_t, E_t, Y_t, Z_t)_{t \in [0, \tau] }$ of \eqref{general single period fbsde with p}-\eqref{relaxed terminal condition with p} started at time $t_0$ and with terminal condition $\Phi(\cdot, \cdot, e_0)$. Explicitly, the relaxed terminal condition \eqref{relaxed terminal condition with p} in this setting is
\begin{align}
\label{relaxed terminal condition with parameter}
\mathbb{P} \left[ \Phi_{-}(P_{\tau} ,E_{\tau}, e_0) \leq \lim_{t \uparrow \tau}Y_t \leq \Phi_{+}(P_{\tau}, E_{\tau}, e_0) \right] = 1.
\end{align}
Now, set $\bar{E}_t = E_t - e_0$ for every $t \in [0,\tau)$. Then, $(P_t, \bar{E}_t, Y_t, Z_t)_{t \in [t_0, \tau] }$ satisfies the same dynamics \eqref{general single period fbsde with p} with $(P_{t_0}, \bar{E}_{t_0}) = (p, 0)$. Moreover, \eqref{relaxed terminal condition with parameter} leads to
\begin{align}
\label{relaxed terminal condition with phi bar and parameter}
\mathbb{P} \left[ \bar{\Phi}_{-}(P_{\tau} , \bar{E}_{\tau}, e_0) \leq \lim_{t \uparrow \tau} Y_t \leq \bar{\Phi}_{+}(P_{\tau}, \bar{E}_{\tau}, e_0) \right] = 1,
\end{align}
where
\begin{align}
\bar{\Phi}(p, \bar{e}, \mathfrak{e}) := \Phi(p, \bar{e} + \mathfrak{e}, \mathfrak{e}),
\end{align}
for every $p \in \R^d$ and $\bar{e}, \mathfrak{e} \in \R$. Notice that $\bar{\Phi}$ belongs to $\hat{\cK}$. Consequently, for the value function, $v^{\bar{\Phi}}$ associated to the terminal condition $\bar{\Phi}$, we must have, by uniqueness, that
\begin{align}
\begin{aligned}
\label{v phi and v phi bar}
Y_0 &= v^{\Phi}(t_0, p, e_0, e_0) \\
 &= v^{\bar{\Phi}}(t_0, p, 0, e_0) .
 \end{aligned}
\end{align}
For each fixed $p \in \R^{d}$ and $\bar{e} \in \R$, the real function $\mathfrak{e}  \mapsto \bar{\Phi}(p, \bar{e}, \mathfrak{e})$ is monotone increasing because $\Phi \in \hat{\cK}$. Therefore, viewing $\mathfrak{e}$ as a parameter, we can apply the comparison property, Proposition \ref{pr some properties 1} (\ref{it comparaison}) along with \eqref{v phi and v phi bar} to show that the first part of the lemma holds; $v^{\Phi}$ has the stated monotonicity property.

2. For the second part, we use the inequality \eqref{mollification proposition inequalities v phi} which is stated in the appendix as part of Proposition \ref{v existence proposition from thesis}. For any $\mathfrak{e} \in \R$, the function $(p,e) \mapsto \bar{\Phi}(p, e, \mathfrak{e})$ belongs to $K$ and so, by \eqref{mollification proposition inequalities v phi} we have, for any $\Lambda > 0$,
\begin{align}
\label{v phi bar inequality in multi period lemma}
v^{\bar{\Phi}}(t_0, p, 0; \mathfrak{e}) \leq e^{-r( \tau -t_0)} \left[ \mathbb{E} \left[ \bar{\Phi}(P_{\tau}, \Lambda, e_0) \right] + C \left( \frac{L( \tau - t_0)}{ \Lambda } \right) \right],
\end{align}	
where $C$ depends on $L$, $L_{\Phi}$, $\tau$ and $p$ only. Now, noting that $\bar{\Phi}(P_{\tau}, \Lambda, \mathfrak{e})$ converges to $0$ as $\mathfrak{e}$ tends to $- \infty$, we take limits in \eqref{v phi bar inequality in multi period lemma}, and use the bounded convergence theorem to give
\begin{align}
\lim_{\mathfrak{e} \rightarrow - \infty} v^{\bar{\Phi}}(t_0, p, 0, \mathfrak{e}) \leq C e^{-r( \tau -t_0)}  \left( \frac{L( \tau - t_0)}{ \Lambda } \right).
\end{align}
Since this holds for any $\Lambda > 0$, we conclude that $\lim_{\mathfrak{e} \downarrow - \infty} v^{\bar{\Phi}}(t_0, p, 0; \mathfrak{e}) = 0$. Combining this observation with \eqref{v phi and v phi bar} completes the proof.
\eproof
\\
\\
The following corollary is useful in the next section as it allows us to link two trading periods.
\begin{Corollary}
\label{le monotonicity for induction corollary}
Using the notation of Lemma \ref{le monotonicity for induction}, for $\Lambda \in \Theta$ and a given $t_0 \in [0,\tau)$, consider the function $\psi$ defined by
\begin{align}
\psi(p,e, \mathfrak{e} ) = \begin{dcases}
v^{\Phi}(t_0,p,e,e) & \text{if } e < \Lambda( \mathfrak{e} ), \\
1 & \text{otherwise.}
\end{dcases}  
\end{align}
Then $\psi \in \hat{\cK}$.
\end{Corollary}
\proof
For the first part of the definition of $\hat{\cK}$, fix a value $\mathfrak{e} \in \R$. By Corollary \ref{co one period general}, for each fixed $e \in \R$, the function $p \mapsto v^{\Phi}(t_0,p,e,e)$ is $C$-Lipschitz continuous for a constant $C$ depending only on $L$, $\tau$ and $L_{\Phi}$, the Lipschitz constant of $\Phi$. {Note that this $C$ is independent of $e$ and the parameter, $\mathfrak{e}$.} Using Lemma \ref{le monotonicity for induction} and the definition of $\psi$, we also see that $(p,e) \mapsto \psi (p,e, \mathfrak{e})$ is monotone increasing in $e$, has limit $0$ as $e$ tends to $- \infty$ and limit $1$ as $e$ tends to $+ \infty$. 

Next, for the second part of the definition of $\hat{\cK}$, fix a value of $e \in \R$. For any $p \in \R^d$ and $\mathfrak{e} \in \R$, we will have
\begin{align}
\psi(p,e + \mathfrak{e}, \mathfrak{e}) = \begin{dcases}
v^{\Phi}(t_0, p, e + \mathfrak{e}, e + \mathfrak{e}) & \text{if } e < \Gamma( \mathfrak{e} ), \\
1 & \text{otherwise,}
\end{dcases}  
\end{align}
where, as in Definition \ref{cap functions set definition}, $\Gamma( \mathfrak{e} ) = \Lambda( \mathfrak{e} ) - \mathfrak{e}$ and $\Gamma$ is a monotone decreasing function such that $\lim_{\mathfrak{e} \rightarrow + \infty} \Gamma( \mathfrak{e} ) = -\infty$. Now, using the same arguments as above, we see that, for each $e \in \R$, the function $(p, \mathfrak{e} ) \mapsto \psi(p, e + \mathfrak{e}, \mathfrak{e})$ is in $K$, recall Definition \ref{de cK}. 
\eproof

\subsection{The multi period model}
The aim of this section is to prove Theorem \ref{multi period model main theorem} and some key properties of the multi-period model that will be used for the infinite period model.

The proof of Theorem \ref{multi period model main theorem} is based on the existence of a pricing function, which is given essentially by the following proposition.

\begin{Proposition}\label{pr pricing function multi-period}
Let $\theta \in \widehat{\cK}$ with Lipschitz constant $L_{\theta}$, and $q$ be a positive integer denoting the number of periods. Let $\Lambda_1$, $\Lambda_2$,...,$\Lambda_{q-1}$ be cap functions in $\Theta$ and $0 < T_1 < T_2 < ... < T_q = T$ be a sequence of times.

There exists a bounded, measurable function $v^q: [0,T) \times \mathbb{R}^d \times \mathbb{R} \times \mathbb{R} \rightarrow \R$ such that
	\begin{enumerate}
		\item{\label{v multi continuity} For each $\mathfrak{e} \in \R$, $v^q(\cdot,\cdot,\cdot,\mathfrak{e} )$  is continuous on $[T_{k-1},T_k)\times \mathbb{R}^d \times \mathbb{R}$, $1 \le k \le q$.
		}
		\item{\label{v multi lipschitz constant e}For any $k$ such that $1 \leq k \leq q$, for any $t \in [T_{k-1},T_k)$, the function $v^q(t,\cdot, \cdot, \cdot)$ is $1/(\ell_1(T_k-t))$-Lipschitz continuous with respect to $e$.}
		\item{\label{v multi lipschitz constant p}For any $t \in  [T_{k-1},T_k)$, the function $v^q(t,\cdot, \cdot,\cdot)$ is $C^{q}_k$-Lipschitz continuous with respect to $p$, where, for each $k$, $C^{q}_k$ is a constant depending on $L$, $L_{\theta}$, $T$ and $q$ only.}
		\item{\label{v multi condition terminal middle} 
			for $1 \le k < q$, define,
			\begin{align} \label{eq term cond middle in corollary}
			\Phi^{q,k}(p, e, \mathfrak{e}) = \begin{dcases}
			v^q(T_{k},p,e,e), & \text{if } e < { \Lambda}_{k}( \mathfrak{e} ) \\
			1, & \text{otherwise,}
			\end{dcases}
			\end{align}
			and set 
			\begin{align}
			\label{eq term cond end}
			\Phi^{q,q}(p,e,\mathfrak{e})= \theta(p,e, \mathfrak{e}),
			\end{align}
			for every $(p,e ) \in \R^d \times \R$ and $\mathfrak{e} \in \R$. Then, for any integer $k$ with $1 \leq k \leq q$ and any $(p,e) \in \R^d \times \R$, we will have
			\begin{align} 
			\Phi^{q,k}_-(p,e,\mathfrak{e}) \le \liminf_{t \uparrow T_k}v^q(t,p_t,e_t,\mathfrak{e}) \le \limsup_{t \uparrow T_k}v^q(t,p_t,e_t,\mathfrak{e}) \le \Phi^{q,k}_+(p,e,\mathfrak{e}),
			\end{align}
for any family $(p_t, e_t)_{0 \leq t < T_k}$ converging to $(p,e)$ as $t \uparrow T_k$.
		}
		
		\item{\label{fbsde solution property in proposition}\sloppy For $1 \le k \le q$, and any  parameters $(t_0,p,e,\mathfrak{e}) \in [T_{k-1},T_k)\times \mathbb{R}^{d} \times \mathbb{R}\times \mathbb{R}$, we have
			$v^q(t,P^{t_0,p}_t,E^{t_0,p,e,\mathfrak{e}}_t,\mathfrak{e}) = Y^{t_0,p,e,\mathfrak{e}}_t$ for $t_0\le t<T_k$ where $(P^{t_0,p},E^{t_0,p,e,\mathfrak{e}},Y^{t_0,p,e,\mathfrak{e}})$ is solution to \eqref{general single period fbsde with p}-\eqref{relaxed terminal condition with p} over the time interval $[t_0, T_k]$ with $(P^{t_0,p}_{t_0},E^{t_0,p,e,\mathfrak{e}}_{t_0}) = (p,e)$  and terminal condition $\Phi^{q,k}(\cdot, \cdot, \mathfrak{e})$.}
	\end{enumerate}
\end{Proposition}
\begin{Remark}
\begin{enumerate}
	\item In general, each of the constants $C^{q}_k$ in part (\ref{v multi lipschitz constant p}) of Proposition \ref{pr pricing function multi-period} depends on the time period $k$ and the number of periods $q$. For each $q$, we will generally have $C^{q}_1 \geq C^{q}_2 \geq .... \geq C^{q}_q = C$, where $C$ is the Lipschitz constant for a one period model on $[T_{q-1}, T]$ with terminal condition $\theta$, as described in part (\ref{v lipschitz constant p in corollary}) of Corollary \ref{co one period general}. Controlling the constants $C^{q}_k$ is key to studying the infinite period model, as we shall see in the next section.
	\item Using standard arguments for FBSDEs, one can show that item (\ref{fbsde solution property in proposition}) of Proposition \ref{pr pricing function multi-period} holds when $p$ and $e$ are replaced by a pair of square integrable $\mathcal{F}_{t_0}$ random variables of the appropriate dimension. The same holds true for part (\ref{fbsde solution property in corollary}) in Corollary \ref{co multi-period contant cap function} below.
\end{enumerate}
\end{Remark}
\proof 
Throughout this proof, we fix the number of periods, $q$ and denote the resulting function by $v$ instead of $v^{q}$. We use induction to show that $v$ has the stated properties on each period $[T_{k-1}, T_k]$, for $1 \leq k \leq q$.
	
\noindent 1. On the last interval, the terminal condition is simply given by $\theta$, which is in $\widehat{\cK}$. Using Corollary \ref{co one period general}, one obtains a function $u$ defined on the time interval $[0,T)$ for the terminal condition $\theta$. We simply set $v(t,p, e, \mathfrak{e}) = u(t,p, e, \mathfrak{e})$ for every $t \in [T_{q-1}, T)$, $p \in \R^d$ and $e, \mathfrak{e} \in \R$, to define $v$ on the last period. All of the properties of $v(t, \cdot, \cdot, \cdot)$ for $t \in [T_{q-1}, T)$ follow from Corollary \ref{co one period general}. Finally, we apply Corollary \ref{le monotonicity for induction corollary} to see that $\Phi^{q, q-1} \in \widehat{\cK}$.\\
2. Induction step. Assume that, for some $k$ with $1 < k < q$, we have that $\Phi^{q, k} \in \hat{\cK}$ and that $v(t, \cdot, \cdot, \cdot)$ has been defined for $t \in [T_k, T)$ and satisfies the properties in the statement of the proposition. Using Corollary \ref{co one period general} with terminal condition $\Phi^{q, k}$ and terminal time $T_k$, one obtains a function $u^{q,k}: [0, T_k) \times \R^d \times \R \times \R$ satisfying the properties stated there. We set $v(t,p, e, \mathfrak{e}) = u^{q,k}(t,p, e, \mathfrak{e})$ for every $t \in [T_{k-1}, T_k)$, $p \in \R^d$ and $e, \mathfrak{e} \in \R$, to define $v$ on the time period $[T_{k-1}, T_k)$. The properties of $v(t, \cdot, \cdot, \cdot)$ for $t \in [T_{k-1}, T_k)$ follow directly from the corresponding properties of $u^{q,k}$ in Corollary \ref{co one period general}. Lastly, we apply Corollary \ref{le monotonicity for induction corollary} to see that $\Phi^{q, k-1} \in \widehat{\cK}$.
\eproof
\color{black}
\subsubsection{Proof of Theorem \ref{multi period model main theorem}} We now turn to the proof of the main result for the multi-period setting and which has been announced in Section \ref{preliminaries}.

\vspace{2mm}
Fix a value of $q$, a starting point $(p_0, e_0)$ and set $v = v^{q}$, the function from Proposition \ref{pr pricing function multi-period}.
\\
\noindent 1. Existence part: Firstly, by classical SDE results \cite{karatzas2012brownian} we have existence of a process, $P$, on $[0,T]$ as defined in \eqref{general single period fbsde without p}. To build $E$ and $Y$, we use then a forward induction. On $[T_0,T_1]$, the first interval, consider $Y^{0,p_0,e_0,e_0}$ given by Proposition \ref{pr pricing function multi-period} above. In particular $E_{T_1}$ is well defined. \\
On $[T_1,T_2]$, we can define, for $t < T_2$,
\begin{align*}
E^{\xi}_t = E_{T_1} + \int_{T_1}^t \mu(P_s,v(s,P_s,E^\xi_s,\xi)) \ud s
\end{align*}
for any (bounded) $\xi$ that is $\cF_{T_1}$ measurable. Indeed, the random drift coefficient 
$e \mapsto \mu(P_s,v(s,P_s,e,\xi))$ is Lipschitz continuous on $[T_1,t]$, $t<T_2$ and adapted. $E$ is thus well defined on $[T_1,T_2)$ and has a limit at $T_2$ by the Cauchy criterion. Let now, $Y^\xi_t = v(t,P_t,E^\xi_t,\xi)$ and observe that $e^{-rt} Y^\xi_t$ is a martingale. Indeed, consider $\xi_n$ a simple random variable with $\xi_n \rightarrow \xi$. Then $e^{-rt} Y^{\xi_n}_t$ is a martingale from the properties of $v^q$ in Proposition \ref{pr pricing function multi-period}. To obtain the result, one can then pass to the limit  and apply the dominated convergence theorem (the processes $Y^{\xi_n}$ are uniformly bounded as $v$ is bounded). Setting, in particular, $\xi = E_{T_1}$ here, we obtain a process $E$ and the corresponding process $Y$ both on $[T_0,T_2)$. One can also verify that the condition \eqref{multi period relaxed terminal condition in theorem} which links the two periods is satisfied, thanks to the properties of the function $v^{q}$ in Proposition \ref{pr pricing function multi-period}.
\\
2. Uniqueness. First, notice that uniqueness of the $P$ process follows by classical results for SDEs driven by Lipschitz continuous coefficients. Now, for $t \in [0, T_1)$, the tuple of processes $(P, E, Y, Z)$ has been constructed so that it is the solution of FBSDE \eqref{general single period fbsde with p}-\eqref{relaxed terminal condition with p} with starting point $(p_0, e_0)$ and terminal condition $\Phi^{q,1}(\cdot, \cdot, e_0)$. Using the same computation as that used to prove uniqueness of solutions in Theorem 2.2 in \cite{carmona2013singularconservation} (see also Theorem \ref{single period main theorem from thesis} in the appendix), we will find that $(E_t, Y_t, Z_t)_{t \in [0,T_1]}$ is unique in $\mathcal{S}^{2,1}_{\mathrm{c}}([0,T_1]) \times \mathcal{S}^{2,1}_{\mathrm{c}}([0,T_1)) \times \mathcal{H}^{2,d}([0,T_1])$. Indeed, this shows that existence and uniqueness in Theorem \ref{multi period model main theorem} holds when $q=1$. Similarly, for any $t_0, T^{\prime} \in [0, T]$ with $t_0 < T^{\prime}$, using Corollary \ref{co one period general}, we obtain existence and uniqueness for processes $(P_t, E_t, Y_t, Z_t)_{t \in [t_0, T^{\prime}]}$ satisfying \eqref{general single period fbsde with p}-\eqref{relaxed terminal condition with p} with a given starting point $(P_{t_0}, E_{t_0})$ and a terminal condition $\Phi \in \hat{\cK}$ and the relevant relaxed terminal condition for $Y_{T^{\prime}}$. Using standard FBSDE arguments and the value function $v^{\Phi}$ in Corollary \ref{co one period general}, we see that the same result holds when $p$, $e$ and $\mathfrak{e}$ are replaced by $\mathcal{F}_{t_0}$ measurable and square integrable random variables; in that case we will still have $Y_t = v^{\Phi}(t, P_t, E_t)$ for $t \in [t_0, T^{\prime})$.

Now, proceeding to the time period $[T_1, T_2]$, we see that for $t \in [T_1, T_2]$, the tuple $(P_t, E_t, Y_t, Z_t)$ has been constructed to be the solution to \eqref{general single period fbsde with p}-\eqref{relaxed terminal condition with p} with the $\mathcal{F}_{T_1}$ measurable starting point $(P_{T_1}, E_{T_1})$ and terminal condition $\Phi^{q, 2}(\cdot, \cdot, E_{T_1})$. By the above argument, existence and uniqueness is seen to hold on the time period $[T_1, T_2]$. 
The proof is completed by an easy induction.
\subsection{Properties of the multi-period model with constant cap functions}
We collect here some key properties of the multi-period model that will be used in the next section. We restrict to a non-parametrised version of the multi-period model, assuming that the cap function are constant. More precisely, we will work throughout this section with Assumption \ref{ass cap infinite model}.

The following results are deduced directly from Theorem \ref{multi period model main theorem} and Proposition  \ref{pr pricing function multi-period} in this restricted setting. They are stated here for the reader's convenience.

\begin{Corollary}\label{co multi-period contant cap function} Let Assumptions \ref{ass coef} and \ref{ass cap infinite model} hold. Let $\theta \in \cK$ with Lipschitz constant $L_{\theta}$, and $q$ be a positive integer denoting the number of periods. We consider a $q$ period multi period model with caps $\Lambda_1$, $\Lambda_2$,...,$\Lambda_{q-1}$ and terminal condition $\theta$ in the final period, over $[0,T_q]$ and time intervals $[T_0, T_1)$, $[T_1, T_2)$,...,$[T_{q-1}, T_q)$.
	
There exists a bounded measurable function $v^q: [0,T) \times \mathbb{R}^d \times \mathbb{R} \rightarrow \R$ such that
\begin{enumerate}
	\item{\label{v multi continuity constant cap} For each $1 \le k \le q$, the function $(t,p,e) \mapsto v^q(t,p,e )$  is continuous on $[T_{k-1},T_k)\times \mathbb{R}^d \times \mathbb{R}$, $1 \le k \le q$;
	}
	\item{\label{v multi lipschitz constant e constant cap}For any $t \in [T_{k-1},T_k)$, the function $v^q(t,\cdot, \cdot)$ is $1/(l_1(T_k-t))$-Lipschitz continuous with respect to $e$.}
	\item{\label{v multi lipschitz constant p constant cap}For any $t \in  [T_{k-1},T_k)$, the function $v^q(t,\cdot, \cdot)$ is $C^{q}_k$-Lipschitz continuous with respect to $p$, where, for each $k$, $C^{q}_k$ is a constant depending on $L$, $L_{\theta}$, $T$ and $q$ only.}
	\item{\label{v multi condition terminal middle constant cap} for $1 \le k < q$, $(e,\mathfrak{e}) \in \R\times\R$, denoting
		\begin{align} \label{eq term cond middle}
		\Phi^{q,k}(p, e) = \begin{dcases}
		v^q(T_{k},p,e), & \text{if } e <  \Lambda_{k} \\
		1, & \text{otherwise.}
		\end{dcases}
		\end{align}
		and setting 
		\begin{align}
		\Phi^{q,q}(p,e)= \theta(p,e),
		\end{align}
		for every $(p,e ) \in \R^d \times \R$,
		we have, given $(p,e) \in \R^d \times \R$ and $k$ such that $1 \leq k \leq q$, for any family $(p_t, e_t)_{0 \leq t < T_k}$ converging to $(p,e)$ as $t \uparrow T_k$, we have,
		\begin{align}
		\Phi^{q,k}_-(p,e) \le \liminf_{t \uparrow T_k}v^q(t,p_t,e_t) \le \limsup_{t \uparrow T_k}v^q(t,p_t,e_t) \le \Phi^{q,k}_+(p,e).
		\end{align}
	}
	
	\item{\label{fbsde solution property in corollary}\sloppy For $1 \le k \le q$, any parameters $(t_0,p,e) \in [T_{k-1},T_k)\times \mathbb{R}^{d} \times \mathbb{R}\times \mathbb{R}$, 
		$v^q(t,P^{t_0,p}_t,E^{t_0,p,e}_t) = Y^{t_0,p,e}_t$ for $t_0\le t< T_k$ where $(P^{t_0,p},E^{t_0,p,e},Y^{t_0,p,e})$ is solution to \eqref{general single period fbsde with p}-\eqref{relaxed terminal condition with p} over the time interval $[t_0, T_k]$ with $(P^{t_0,p}_{t_0},E^{t_0,p,e}_{t_0}) = (p,e)$ and terminal condition $\Phi^{q,k}$. 
	}
\end{enumerate}
Finally, assume that we are given two deterministic starting points $P_0 \in \mathbb{R}^d$, $E_0 \in \mathbb{R}$. Then there exists a unique c\`{a}dl\`{a}g process $(Y_t)_{0 \leq t \leq T} \in \mathcal{S}^{2,1}([0,T])$, a continuous process $(E_t)_{0 \leq t \leq T} \in \mathcal{S}^{2,1}_{c}([0,T])$ and a continuous process $(P_t)_{0 \leq t \leq T} \in \mathcal{S}^{2,1}_{c}([0,T])$  satisfying the dynamics \eqref{general single period fbsde without p} on each period $[T_{k-1},T_k)$, $1 \leq k \leq q$. For each $k$, the process $Y$ is  continuous on $[T_{k-1},T_k)$; it can have a jump at $T_k$. There it satisfies, for every $1 \leq k \leq q$, almost surely,
\begin{align}
\begin{aligned}
\label{multi period relaxed terminal condition constant caps}
\lim_{t \uparrow T_k} Y_t &= Y_{T_k},  && \text{if } E_{T_k} < {\Lambda}_k, \\
\lim_{t \uparrow T_k} Y_t &= 1, && \text{if } E_{T_k} > {\Lambda}_k, \\
Y_{T_k} \leq &\lim_{t \uparrow T_k} Y_t \leq 1, && \text{if } E_{T_k} = {\Lambda}_k, 
\end{aligned}
\end{align}
with, as above, $Y_{T_q} = 0$.
\end{Corollary}
\vspace{2mm}
From now on, unless stated otherwise, for $\theta$ in Proposition \ref{pr pricing function multi-period}, we always set
\begin{align}
\label{theta final period non constant cap}
\theta(p,e, \mathfrak{e}) = \mathbf{1}_{ e \geq \Lambda_q (\mathfrak{e})  }, \quad (p,e) \in \R^d \times \R,
\end{align}
and similarly for constant caps and the setting of Corollary \ref{co multi-period contant cap function}:
\begin{align}
\label{theta definition constant cap}
\theta(p,e) = \mathbf{1}_{ e \geq \Lambda_q }. 
\end{align}
This is the standard setting of a multi period model as in Theorem \ref{multi period model main theorem}. Indeed, for a single period model for a carbon market, the standard terminal condition is, when the cap on emissions is $\Lambda$ and penalty for over emission is $\pi$ > 0, given by $\phi(e) = \pi \mathbf{1}_{ e \geq \Lambda}$. Terminal conditions of this form are standard for singular FBSDEs modelling carbon markets in the literature; see \cite{carmona2012valuation, carmona2013singular, carmona2009optimal, carmona2010market,carmona2011risk, chesney2012endogenous, howison2012risk} among others. 

The final period of a multi period model is no different to a single period model and so, taking into account the variable caps and the fact that we set the penalty equal to $1$ throughout this paper, \eqref{theta final period non constant cap} is the corresponding terminal condition to be used for the final period.

\begin{Lemma} \label{le prog dyn}
Let Assumptions \ref{ass coef} and \ref{ass cap infinite model} hold. For any positive integer $q$, denote by $v^q$ the function defined in Corollary \ref{co multi-period contant cap function} for $\theta$ given by \eqref{theta definition constant cap}. Then, we have, for $q \ge 2$
\begin{align}
\label{v time emissions period translation}
v^q(T_k,p,e) = v^{q-1}(T_{k-1},p,e-\lambda),
\end{align}
for every integer $1 \le k \le q-1$ and $(p,e) \in \R^d \times \R$.
\end{Lemma}

\proof
1. Let $x \in \R$ be fixed. For $\phi \in \cK$, we define $\tilde{\phi}(p,e) := \phi(p,e-x)$, for $(p,e) \in \R^d\times \R$. Note that $\tilde{\phi} \in \cK$.
Let $v$ (respectively $\tilde{v}$) be the function given in Proposition \ref{pr existence uniqueness one-period} and associated to $\phi$ (respectively $\tilde{\phi}$). The goal of this step is to show that
\begin{align} \label{eq invariance one period}
v(t,p,e-x) = \tilde{v}(t,p,e) \text{ for } (t,p,e) \in [0,\tau)\times \R^d \times \R\,.
\end{align}
For a $t_0 < \tau$, let $(P^{t_0,p}_t)_{t_0 \le t \le \tau}$ be the solution of \eqref{multi period P equation} satisfying $P^{t_0,p}_{t_0} = p$. Denote by $(P^{t_0,p}_t, \tilde{E}^{t_0,p,e}_t, \tilde{Y}^{t_0,p,e}_t)$ the $(P,E,Y)$ part of the unique solution of \eqref{general single period fbsde with p} given in Proposition \ref{pr existence uniqueness one-period} over time interval $[t_0, \tau]$ with $(P^{t_0,p}_{t_0}, \tilde{E}^{t_0,p,e}_{t_0}) = (p,e)$ and with terminal condition $\tilde{\phi}$. Similarly, let $(P^{t_0,p}_t, E^{t_0,p,e}_t, Y^{t_0,p,e}_t)$ be the $(P,E,Y)$ part of the solution of \eqref{general single period fbsde with p} with the same initial condition, over the same time interval, and with terminal condition $\phi$. For the first solution tuple, the following relaxed terminal condition is satisfied. 
\begin{align}
\mathbb{P} \left[ \tilde{\phi}_{-}(P^{t_0,p}_{\tau},\tilde{E}^{t_0,p,e}_{\tau}) \leq \tilde{Y}^{t_0,p,e}_{\tau -} \leq \tilde{\phi}_{+}(P^{t_0,p}_{\tau}, \tilde{E}^{t_0,p,e}_{\tau}) \right] = 1.
\end{align}
We observe that the previous condition rewrites
\begin{align}
\mathbb{P} \left[ {\phi}_{-}(P^{t_0,p}_{\tau},\tilde{E}^{t_0,p,e}_{\tau}-x) \leq \tilde{Y}^{t_0,p,e}_{\tau} \leq {\phi}_{+}(P^{t_0,p}_{\tau}, \tilde{E}^{t_0,p,e}_{\tau}-x) \right] = 1.
\end{align}
by definition of $\tilde{\phi}$. By the uniqueness result of Proposition \ref{pr existence uniqueness one-period}, we observe then that 
\begin{align*}
(P^{t_0,p}_t, \tilde{E}^{t_0,p,e}_t - x, \tilde{Y}^{t_0,p,e}_t) = (P^{t_0,p}_t, {E}^{t_0,p,e-x}_t, {Y}^{t_0,p,e-x}_t)\, t_0 \le t < \tau,
\end{align*}
where $({E}^{t_0,p,e-x}, {Y}^{t_0,p,e-x})$ are parts of the solution $(P^{t_0,p},E^{t_0,p,e-x},Y^{t_0,p,e-x},Z^{t_0,p,e-x})$, of \eqref{general single period fbsde with p}, over $[t_0, \tau]$ with initial condition $(P^{t_0,p}_{t_0},E^{t_0,p,e-x}_{t_0}) = (p, e - x)$ and  terminal condition $\phi$, as they have the same dynamics and relaxed terminal condition.
In particular, we obtain \eqref{eq invariance one period} for $t = t_0$.
\\
2. Now, for any $\phi \in \cK$ and $T > 0$, we will denote here by $u^{\phi, T}$ the value function $v$ from Proposition \ref{pr existence uniqueness one-period} for the system \eqref{general single period fbsde with p} over $[0, T]$ and with terminal condition $\phi$. For this step, we note that for any $\phi \in \cK$, $T > 0$ and $\Delta T > 0$, we have
\begin{align}
\label{value function time invariance}
u^{\phi, T} (t, p, e) = u^{\phi, T + \Delta T} (t + \Delta T, p, e),
\end{align}
for every $t \in [0, T)$ and every $(p, e) \in \R^d \times \R$. Indeed, this property is easily seen to hold for functions $v^{\epsilon, n}$ satisfying the PDE system \eqref{value function pde}. Approximating $\phi$ by a sequence of smooth functions and using Corollary \ref{le smooth approx of v}, we conclude that \eqref{value function time invariance} holds.

\noindent 3. In this step, we prove \eqref{v time emissions period translation} by induction on the period number, $k$.

\noindent 3a. Induction step. For a $k$ such that $2 \le k \le q-1$, assume that \eqref{v time emissions period translation} holds. In this setting, this leads to
\begin{align}\label{eq induction hypothesis invariance}
\Phi^{q-1,k-1}(p,e-\lambda) = \Phi^{q,k}(p,e)\;,
\end{align}
where the terminal condition functions $\Phi^{\cdot, \cdot}$ are as in Corollary \eqref{co multi-period contant cap function}.

By construction, recall the proof of Proposition \ref{pr pricing function multi-period}, $v^q$ is given on  $[T_{k-1}, T_k)$ by $u^{\Phi^{q,k}, T_{k}}$ over the same time interval. That is,
\begin{align}
v^q(t,p,e) = u^{\Phi^{q,k}, T_{k}}(t,p,e),
\end{align}
for every $t \in [T_{k-1},T_{k})$ and $(p,e) \in \R^d \times \R$. Similarly, $v^{q-1}$ on the time interval $[T_{k-2},T_{k-1})$ matches $u^{\Phi^{q-1,k-1}, T_{k-1}}$ on the same time interval. Using step 2 followed by step 1, we compute
\begin{align*}
v^{q}(T_{k-1}, p, e) &= u^{\Phi^{q,k}, T_{k}}(T_{k-1},p,e) \\ &= u^{\Phi^{q,k}, T_{k-1}}(T_{k-2},p,e) \\
&= u^{\Phi^{q-1,k-1}, T_{k-2}}(T_{k-2},p,e - \lambda ) \\
&= v^{q - 1}(T_{k-2}, p, e - \lambda),
\end{align*}
for every $(p,e) \in \R^d \times \R$, completing the induction step.

\noindent 3b. For the initialization step, we note that
\begin{align*}
\Phi^{q,q}(p,e) := \1_{e \ge \Lambda_q} = \1_{e-\lambda \ge \Lambda_{q-1} } = \Phi^{q-1,q-1}(p,e-\lambda)\;.
\end{align*}
Using the same computation as in part 3a, this leads to \eqref{v time emissions period translation} for $k=q-1$. 
\eproof

\begin{Remark} \label{re picard iteration = multi period}
Focusing on $v^q$ in the first period $[0,\tau)$, we observe that it can be computed by the following iteration:
\begin{enumerate}[i)]
\item for $n=0$, set $\omega^0 = 0$ on $[0,\tau)$.
\item for $n\ge 1$, $\omega^n$ is obtained by solving the one-period FBSDE with terminal condition given by
\begin{align} \label{eq term cond middle picard}
        \Phi^n:(p,e) = \begin{dcases}
       \omega^{n-1}(0,p,e-\lambda), & \text{if } e <  \lambda, \\
        1, & \text{otherwise,}
        \end{dcases}
\end{align}
using Proposition \ref{pr existence uniqueness one-period}.
\end{enumerate}
Then, it follows from Lemma \ref{le prog dyn} above, that, for $q \ge 1$, $t \in [0,\tau)$ and $(p,e) \in \R^d\times \R$, we have $v^q(t,p,e)=\omega^q(t,p,e)$.
\end{Remark}

It is then completely natural to ask if the functions $\omega^q$ converges when $q \rightarrow \infty$. A first result in this direction is the following.

\begin{Proposition}\label{pr definition of w}
The sequence $(\omega^q)_{q \ge 0}$ is bounded and increasing, in the sense that, for $q \ge 1$,
\begin{align*}
\omega^q(t,p,e) \ge \omega^{q-1}(t,p,e)\;,\quad \forall (t,p,e) \in [0,\tau)\times \R^d \times \R\;.
\end{align*}
Their limit $w$ is well defined. Moreover, if $r > 0$, it satisfies, for all $t <  \tau$,
\begin{align}\label{eq L1 integrability of w}
\sup_{p \in \R^d}\int_{-\infty}^0 w(t,p,e) \ud e < +\infty\;.
\end{align}
\end{Proposition}

\proof 1. The monotonicity of the sequence is proved by induction. Assume that, for some $n\ge 1$, that $w^{n-1} (t,p,e) \leq w^{n} (t,p,e)$ for every $(t,p,e) \in [0, \tau) \times \R^d \times \R$. Then $\Phi^{n-1}(p,e) \leq \Phi^{n}(p,e)$ for every $(p,e) \in \R^d \times \R$ by  construction, recall \eqref{eq term cond middle picard}. Now, using the comparison principle, Proposition \ref{pr some properties 1}(\ref{it comparaison}), we obtain that $w^{n}(t,p,e) \leq w^{n+1}(t,p,e)$ for every $(t,p,e) \in [0, \tau) \times \R^d \times \R$. Observe that a direct application of the comparison principle leads also to $w^{0}(t,p,e) \leq w^{1}(t,p,e)$ for every $(t,p,e) \in [0, \tau) \times \R^d \times \R$.
\\ 
2. The pointwise convergence comes from the fact that the $(\omega^q)_{q\ge 0}$ are trivially bounded by one.
By monotone convergence, 
$$\lim_{n \rightarrow \infty} \int_{-\infty}^0 \omega^n(0,p,e) \ud e = \int_{-\infty}^0 w(0,p,e) \ud e\,.$$
Now, using \eqref{eq L1 integrability}, we obtain
\begin{align}\label{eq rewrites L1 int omega n}
\sup_{p \in \R^d} \int_{-\infty}^{0}\omega^n(t,p,e) \ud e \le e^{-r(\tau-t)} \left( \sup_{p\in\R^d} \int_{-\infty}^{0}\Phi^n(p,e) \ud e + C \right).
\end{align}
This leads, in particular at $t=0$,
\begin{align*}
\sup_{p \in \R^d} \int_{-\infty}^{0}\omega^n(0,p,e) \ud e \le e^{-r\tau} \left( \sup_{p\in\R^d} \int_{-\infty}^{0}\omega^{n-1}(0,p,e) \ud e + C \right),
\end{align*}
recalling the definition of $\Phi^n$ and the non-negativity of $\omega^{n-1}$. At the limit, we thus obtain \eqref{eq L1 integrability of w} at $t=0$. Combining this with \eqref{eq rewrites L1 int omega n} and a monotone convergence argument concludes the proof.
\eproof

\vspace{2mm}
The goal of the next section is to identify precisely the limit of the sequence of functions $(\omega^q)_{q \ge 0}$ with the definition of the infinite period model.

\section{Study of the Infinite Period Model}
\label{mainresults2}

This section is dedicated to the proof Theorem \ref{th infinite model 1}.
In a first part, we will build the function $w$ and the associated FBSDE. We show that it is the limit of a $q$ period pricing function when $q$ goes to infinity.
In a second part, we will prove uniqueness of this function.

\vspace{2mm}
As usual, in order to build the solution to the FBSDE, we first exhibit a pricing function (decoupling field).

\begin{Proposition}\label{pr existence pricing function infinite model} 
Let Assumptions \ref{ass coef}, \ref{ass cap infinite model} hold. Assume also that Assumption \ref{ass infinite model coef}  or Assumption \ref{ass unif ellip} holds.
\\
There exists a continuous function
  $w:[0,\tau)\times \R^d \times \R \rightarrow \R$
satisfying
\begin{enumerate}  
\item{\label{w lipschitz constant e}For any $t \in [0,\tau)$, the function $w(t,\cdot, \cdot)$ is $1/(\ell_1(\tau-t))$-Lipschitz continuous with respect to $e$,}
		\item{\label{w lipschitz constant p}For any $t \in [0,\tau)$, the function $w(t,\cdot, \cdot)$ is $C$-Lipschitz continuous with respect to $p$, where $C>0$ is a constant depending on $L$, $\tau$ and $\beta$ only.}
		\item{\label{w condition terminal} Given $(p,e) \in \R^d \times \R$, for any family $(p_t, e_t)_{0 \leq t < \tau}$ converging to $(p,e)$ as $t \uparrow \tau$, we have
		\begin{align}\label{eq w condition terminal}
		\Phi^w_-(p,e) \le \liminf_{t \uparrow \tau}w(t,p_t,e_t) \le \limsup_{t \uparrow \tau}w(t,p_t,e_t) \le \Phi^w_+(p,e)
		\end{align}
where
\begin{align}
\label{Theta k definition}
\Phi^w(p,e) = \begin{dcases} w(0, p, e - \lambda), & \text{if }       e < \lambda, \\
1, & \text{otherwise.}
\end{dcases}
\end{align}
}
\item{\label{w integral bound in proposition} 
	The following holds
\begin{align*}
\sup_p \int_{-\infty}^0 w(0,p,e) \ud e < \infty
\end{align*}}
\item {
\label{w martingale and terminal condition in proposition}	
For any $(t_0,p,e)\in [0,\tau)\times \R^d \times \R$, denoting $P^{t_0,p}$ the unique strong solution to \eqref{multi period P equation}, the unique strong solution $(E^{t_0,p,e}_t)_{t_0 \le t < \tau}$ to
\begin{align}
E^{t_0,p,e}_t = e + \int_{t_0}^t\mu(P^{t_0,p}_s,w(s,P^{t_0,p}_s,E^{t_0,p,e}_s)) \ud s, \; t_0 \le t < \tau
\end{align}
is such that $(e^{-rt}Y_t)_{t_0 \le t < \tau}$, with $Y_t=w(t,P^{t_0,p}_t,E^{t_0,p,e}_t)$, is a $[0,1]$-valued $\cF$-martingale. Then, the limit $\lim_{t \uparrow \tau } Y_t$ exists and it satisfies
\begin{align}
\label{Y infinite period terminal condition first period}
\P\left(\,\Phi^w_-(P^{t_0,p}_\tau,E^{t_0,p,e}_\tau)\le \lim_{t \uparrow \tau } Y_t \le \Phi^w_+(P^{t_0,p}_\tau,E^{t_0,p,e}_\tau)\,\right) = 1.
\end{align}}
\end{enumerate}
\end{Proposition}

\proof
Recall that $(\omega^n)_{n \ge 1}$ denotes the sequence introduced in Remark \ref{re picard iteration = multi period} and that $w = \lim_n \omega^n$ is well defined by Proposition \ref{pr definition of w}.
\\
Part (\ref{w lipschitz constant e}): By Proposition \ref{pr existence uniqueness one-period} each function $w^n$ satisfies item \ref{w lipschitz constant e} in the statement of the proposition and, therefore, the same holds for $w$.
\\
Part (\ref{w lipschitz constant p}): Let  $L_n^0$ be the Lipschitz constant in $p$ of $\omega^n(0,\cdot, \cdot)$.
\begin{enumerate}[i)]
\item Under Assumption \ref{ass infinite model coef}:  By Proposition \ref{pr existence uniqueness one-period}, it depends only on $L$, $L_{\Phi^n}$ and $\tau$. Precisely, from Lemma \ref{le control lipschitz}, we have that
\begin{align*}
L^0_n \le \frac{L}{l_1} + e^{- \beta \tau} L^0_{n-1}.
\end{align*}
This leads to $\lim_{n \rightarrow \infty} L^0_n < \infty$.

\noindent Moreover, using Lemma \ref{le control lipschitz} again, we also have that 
\begin{align*}
|\omega^n(t,p,e)-\omega^n(t,p',e)| \le \left( \frac{L}{l_1} + L^0_{n-1} \right) |p-p'|.
\end{align*}
By taking limits we deduce that $w$ is $C$-Lipschitz in $p$ with $C = L/{l_1} +\lim_{n \rightarrow \infty} L^0_n$.
\item Under Assumption \ref{ass unif ellip}: From Lemma \ref{le other control lipschitz}, we have that
\begin{align*}
L^0_n &\le C_{\beta}\left( \frac{\|\Phi^n\|_\infty}{\sqrt{\tau}} +   \frac{1}{l_1}\right) \le \tilde{C}\,,
\end{align*}
recall \eqref{eq term cond middle picard} and $0 \le \omega^{n-1} \le 1$. The constant $\tilde{C}$ depends on $\beta$, 
$L$ and $\tau$ but, importantly, it is uniform in $n$. Now, by Proposition \ref{pr existence uniqueness one-period}, each $\omega^n$ is Lipschitz continuous with a Lipschitz constant depending on $\tilde{C}$ and so is their limit $\omega$.

\end{enumerate}
Part (\ref{w integral bound in proposition}) follows directly from Proposition \ref{pr definition of w}.
\\
Finally, for the functions $\Phi^{n}$ given by \eqref{eq term cond middle picard} we see that $\lim_{n \rightarrow \infty} \Phi^{n} = \Phi^{w}$. Using Proposition \ref{v phi limit proposition more general from appendix} from the appendix, we conclude that $w$ is a value function for the singular FBSDE \eqref{general single period fbsde with p} with terminal condition $\Phi^{w}$. Then, parts (\ref{w condition terminal}) and (\ref{w martingale and terminal condition in proposition}) follow from Proposition \ref{pr existence uniqueness one-period} and the original result, Proposition \ref{v existence proposition from thesis}, in the appendix.
\eproof
\vspace{2mm}
\begin{Proposition} \label{pr uniqueness pricing function infinite model}
Let Assumptions \ref{ass coef} and \ref{ass cap infinite model} hold and assume $r>0$. The function $w$ defined in Proposition \ref{pr existence pricing function infinite model} is unique.
\end{Proposition}

\proof 
Let ${}^1w$ and ${}^2w$ be two continuous functions satisfying the items in the statement of Proposition \ref{pr existence pricing function infinite model}. Set $\Delta w := {}^1w- {}^2w$ and $\Delta \Phi := \Phi^{{}^1w}-\Phi^{{}^2w}$. Using Lemma \ref{le l1 estimate one period}
\begin{align}
\label{eq uniqueness w}
\left(\int |\Delta w|(0,p,e) \ud e\right) & \le e^{-r \tau} \esp{\left(\int |\Delta \Phi|(P^{0,p}_\tau,e)  \ud e\right)} \nonumber
\\
&\le e^{- r \tau}  \sup_p \left(\int |\Delta \Phi|(p,e)  \ud e\right) < \infty 
\end{align}
We have used here that $\Delta \Phi (p,e)=\Delta w(0,p,e-\lambda)\1_{\set{e < \lambda}}$ and \eqref{eq L1 integrability of w}. This, combined with \eqref{eq uniqueness w} leads to
\begin{align*}
(1-e^{- r \tau}) \sup_p \left( \int |\Delta w|(0,p,e) \ud e \right) \le 0.
\end{align*}
By continuity of $\Delta w$, we obtain that ${}^1w(0,p,e)={}^2w(0,p,e)$ for every $(p,e) \in \R^d \times \R$. Then, $\Phi^{{}^1w}$ and $\Phi^{{}^2w}$ are equal, and, ${}^1w$ and ${}^2w$ are value functions, in the sense of Proposition \ref{pr existence uniqueness one-period} for the same FBSDE with the same  terminal condition. By uniqueness of such value functions, which follows from the uniqueness in Proposition \ref{pr existence uniqueness one-period} or by \eqref{eq l1 estimate one period}, we conclude that ${}^1w(t,p,e)={}^2w(t,p,e)$ for every $(p,e) \in \R^d \times \R$ and all $t \in [0, \tau)$.
\eproof

\vspace{2mm}
\noindent We conclude this section by the proving the main result for the infinite period model.

\vspace{2mm}
\noindent \textbf{Proof of Theorem \ref{th infinite model 1}}. 
We build the tuple of processes $(P,E,Y, Z)$ by a forward induction, similarly to the proof of Theorem \ref{multi period model main theorem}.
\\
1.a For the first period, we set $(E_t)_{t \in [0,\tau)}$ to be the solution to
\begin{align*}
E_t = e + \int_0^t\mu(P_s,w(s,P_s,E_s)) \ud s,
\end{align*}
with $P$ being the strong solution to \eqref{multi period P equation} on time interval $[0, \tau]$ with $P_0=p$. By the Cauchy criterion, we can take the limit as $t \uparrow \tau$ to define $E_t$ continuously at $t =\tau$. For $t \in [0,\tau)$, we set $Y_t := w(t,P_t,E_t)$. Notice that, by construction, $w$ is the unique value function described in Propositions \ref{pr existence pricing function infinite model} and \ref{pr uniqueness pricing function infinite model}, and it satisfies all the properties described therein. In particular, the process $(e^{-rt}Y_t)_{0 \leq t < \tau}$ as a $[0,1]$ valued $\mathcal{F}$-martingale. Therefore, for $\hat{Y}_t:= e^{-rt}Y_t$, we see that $\lim_{t \uparrow \tau} \hat{Y}_t$ exists and is well defined. Then, the same holds for $\lim_{t \uparrow \tau} Y_t$ because $Y_t = e^{rt} \hat{Y_t}$. For this process, the terminal condition \eqref{Y infinite period terminal condition first period} will be satisfied. For $t \in [0, \tau)$ we define $(Z_t)_{0 \leq t < \tau}$ such that $(e^{-rt} Z_t){0 \leq t < \tau}$ is the integrand in the martingale representation for $(e^{-rt}Y_t)_{0 \leq t < \tau}$ as a stochastic integral with respect to $W$.
\\
1.b For a positive integer $k$, assume that $(P_t)_{0 \le t \le T_k}$, $(E_t)_{0 \le t \le T_k}$, $(Y_t)_{0 \le t < T_k}$ and $(Z_t)_{0 \le t < T_k}$ are well defined and satisfy the properties described in the statement of Theorem \ref{th infinite model 1}. On $[T_k,T_{k+1})$, we define $E$ as the solution to
\begin{align*}
E_t = E_{T_{k}} + \int_{T_k}^t\mu(P_s,w(s-T_k,P_s,E_s-\Lambda_k)) \ud s \;,\; T_k \le t < T_{k+1}\,
\end{align*}
and define $E_{T_{k+1}}$ continuously, using the Cauchy criterion. Above, $P$ is solution to \eqref{multi period P equation} on $[T_k,T_{k+1}]$ with starting point at $T_k$ equal to $P_{T_k}$, coming from the induction hypothesis. For $t \in [T_k,T_{k+1})$, we simply set $Y_t := w(t-T_k,P_t,E_t-\Lambda_k)$. By Proposition \ref{pr existence pricing function infinite model}, the process $(e^{-rt}Y_t)_{T_{k} \leq t < T_{k+1}}$ is a $[0,1]$ valued $\mathcal{F}$-martingale; we define $Z_t$ for $t \in [T_k,T_{k+1})$ such that $(e^{-rt} Z_t)_{T_{k} \leq t < T_{k+1}}$ is the integrand in the martingale representation for $(e^{-rt}Y_t)_{T_{k} \leq t < T_{k+1}}$ as a stochastic integral with respect to $W$.
\\
By induction, this concludes the proof of the existence of a process $(P,E,Y, Z)$ satisfying \eqref{fbsde time homogeneous}
on each period $[T_k,T_{k+1})$, $k \ge 0$, in lieu of terminal condition.
\\
2. To conclude the proof, we now check \eqref{infinite period relaxed terminal condition in theorem}. Set $k \ge 1$.  For $t \in [T_{k-1},T_k)$, we thus have, by the previous step, that
$Y_t = w(t-T_{k-1},P_t,E_t - \Lambda_{k-1})$. If $E_{T_k} - \Lambda_{k-1} < \lambda$, then by continuity of $(P,E)$ and the properties \eqref{eq w condition terminal} and \eqref{Y infinite period terminal condition first period}, we obtain that $\lim_{t \uparrow T_k} Y_t = w(0,P_{T_k},E_{T_k} - \Lambda_{k-1}-\lambda) = Y_{T_k}$, by construction. 
Similarly, by \eqref{eq w condition terminal} and \eqref{Y infinite period terminal condition first period}, we will find that $\lim_{t \uparrow T_k} Y_t = 1$ when $E_{T_k} > \Lambda_k$, and that $Y_{T_k} \leq \lim_{t \uparrow T_k} Y_t \leq 1$ when $E_{T_k} = \Lambda_k$. This proves \eqref{infinite period relaxed terminal condition in theorem}.
\eproof

\section*{Acknowledgements}
The authors would like extend their gratitude to Dr Mirabelle Mu\^uls, whose collaboration and input greatly influenced the direction and applicability of this research. The authors would also like to thank the Engineering and Physical Sciences Research Council (EPSRC), Climate-KIC and everybody involved with the Centre for Doctoral Training in the Mathematics of Planet Earth at Imperial College London and the University of Reading. The EPSRC and Climate-KIC funded all parts of Hinesh Chotai's PhD programme in the centre for doctoral training, and provided many opportunities for travel to conferences and events during the programme. Dan Crisan would also like to thank \'Ecole Polytechnique and Universit\'e Paris Diderot for its visitor grants. 
\bibliographystyle{plain}
\bibliography{ChChCrbib}
\appendix

\counterwithin{theorem}{section}
\section{}
\label{appendix}
The proofs of some results for the multi period model in Section \ref{mainresults1} rely on some corresponding results for a single period model. A one period model is, as described in Proposition \ref{pr existence uniqueness one-period}, a FBSDE of the form \eqref{general single period fbsde with p} with $\phi \in \cK$ and $b$, $\sigma$, $\mu$ satisfying the conditions in Assumption \ref{ass coef}. This setting is very similar to the setting considered in \cite{carmona2013singularconservation}, Section 2. The main difference is the fact that the terminal condition $\phi$, is a function of two variables, $p$ and $e$ and that $\phi$ is $L_{\phi}$ Lipschitz continuous in the $p$ variable. In contrast, in \cite{carmona2013singularconservation}, $\phi$ was assumed to be a real valued function
\begin{align*}
\begin{aligned}
\phi: \R &\rightarrow [0,1], \quad \\
e &\mapsto \phi(e)
\end{aligned}
\end{align*}
which is monotone increasing and satisfies
\begin{align*}
\lim_{e \downarrow - \infty} \phi(e) &= 0, \\
\lim_{e \uparrow + \infty} \phi(e) &= 1.
\end{align*}
Consequently, the proof of Proposition \ref{pr existence uniqueness one-period} can be proved using very similar arguments as those in Section 2 of \cite{carmona2013singularconservation}.

Proposition \ref{pr existence uniqueness one-period} is a result of the following two results, which are the versions of Theorem 2.2 and Proposition 2.10 from \cite{carmona2013singularconservation} in this setting. The proofs are very similar to the corresponding proofs there and are omitted \cite{chotai2019thesis}.

\begin{proposition}
	\label{v existence proposition from thesis}
	Consider the system \eqref{general single period fbsde with p} with a terminal condition $\phi \in \cK$and set $t_0 = 0$ to simplify notation (the analogous result holds for any $t_0 \in [0, \tau)$). There exists a unique continuous function $v: [0,\tau) \times \mathbb{R}^{d} \times \mathbb{R} \rightarrow [0,1]$, $(t,p,e) \mapsto v(t,p,e)$, called a value function or decoupling field, such that
\begin{enumerate}
	\item{\label{v lipschitz constant e} For any $t \in [0,\tau)$, the function $v(t,\cdot, \cdot)$ is $1/(\ell_1(\tau -t))$-Lipschitz continuous with respect to $e$,}
		\item{\label{v lipschitz constant p}For any $t \in [0,\tau)$, the function $v(t,\cdot, \cdot)$ is $C$-Lipschitz continuous with respect to $p$, where $C$ is a constant depending on $L$, $L_{\phi}$ and $\tau$ only,}
		\item{\sloppy For any initial parameters $(p,e) \in \mathbb{R}^{d} \times \mathbb{R}$, let $(P_t)_{0 \leq t \leq \tau}$ be the unique strong solution of the forward equation for $P$ in \eqref{general single period fbsde with p} with $P_0 = p$. Then, the unique strong solution, $(E_t)_{0 \leq t < \tau}$ of
			\begin{align}
			\label{E t definition}
			E_t = e + \int_{0}^t \mu(P_s, v(s, P_s, E_s) ) \dif s, \quad 0 \leq t < \tau,
			\end{align}
			is such that the process $( e^{-rt} v(s, P_t, E_t) )_{0 \leq t < \tau}$ is a $[0,1]$-valued martingale with respect to the complete filtration generated by $W$.
			
			The limit $\lim_{t \uparrow \tau} E_t$ exists almost surely and so does $\lim_{t \uparrow \tau}( e^{-rt} v(s, P_t, E_t) )$, being the limit of a bounded martingale.}			
			\item{Let $Z$ be such that the process $(e^{-rt} Z_t)_{0 \leq t \leq \tau}$ is the integrand in the martingale representation for $(e^{-rt}v(t, P_t, E_t) )_{0 \leq t \leq \tau}$ as a stochastic integral with respect to $W$. Then $Z$ satisfies the boundedness conditions in Theorem \ref{single period main theorem from thesis} below: if $\sigma$ is bounded by $L$ then there exists a constant $C$ depending only on $L$, $L_{\phi}$ and $\tau$, such that $|Z_t| \leq C$ for almost every $(t, \omega) \in [0,\tau] \times \Omega$. On the other hand, for general $\sigma$ satisfying only the linear growth and Lipschitz continuity conditions in Assumption \ref{delarue assumptions}, then, given an initial value $p \in \R^{d}$ for the forward equation for $P$ in \eqref{general single period fbsde with p}, and any $a \geq 1$, $Z$ satisfies \eqref{Z bound in single period theorem} in Theorem \ref{single period main theorem from thesis} below, for almost every $(t, \omega) \in [0,\tau] \times \Omega$, where $C^{\prime} > 0$ in \eqref{Z bound in single period theorem} is a constant depending only on $L$, $L_{\phi}$, $\tau$ and $a$.}
			\item{\label{v estimate in appendix}Given $\xi>0$ and $a \geq 1$, there exists a constant $C(\xi, a) > 0$ depending only on $a$, $\xi$, $L$ and $\tau$ such that for all $t \in [0, \tau)$, all $e, \Lambda \in \R$ and every $p \in \R^d$ such that $|p| \leq \xi$, we have	
			\begin{align}
			\label{mollification proposition inequalities v phi}
			\begin{aligned}
			e > \Lambda \implies v(t,p, e) \geq e^{-r(\tau-t)} \left( \mathbb{E}[ \phi(P_T^{ t,p}, \Lambda) | ] - C(\xi, a) \left( \frac{e - \Lambda}{L(\tau-t)} \right)^{-a} \right) , \\
			e < \Lambda \implies v(t,p,e) \leq e^{-r( \tau -t)} \left( \mathbb{E}[ \phi(P_T^{ t,p}, \Lambda)] + C(\xi, a) \left( \frac{\Lambda - e}{L(\tau-t)} \right)^{-a} \right).
			\end{aligned}
			\end{align}
		The upper bound can also be given in the following form. Given $a \geq 1$, there is a constant $C > 0$ depending only on $L$ and $\tau$ such that, for every $t \in [0, \tau)$ and $p \in \R^d$ we have
	\begin{align}
		v(t,p,e) \leq e^{-r(\tau-t)} \left( \mathbb{E}[ \phi(P_T^{ t,p}, \Lambda)] + \min \left( C \left( 1 + |p|^a \right) \left( \frac{\Lambda - e}{L(\tau-t)} \right)^{-a} , 1 \right)  \right), 
\end{align}
whenever $e < \Lambda$.}
		\item{For any $t \in [0,\tau)$, $p \in \mathbb{R}^d$ the function $e \mapsto v(t,p,e)$ is monotone increasing and satisfies
			\begin{align}
			\begin{aligned}
			\label{v infinity limits}
			\lim_{e \rightarrow \infty} v(t,p,e) &= e^{-r(\tau-t)}, \\
			\lim_{e \rightarrow - \infty} v(t,p,e) &= 0.
			\end{aligned}
			\end{align}}
	\item{Given $(p,e) \in \R^d \times \R$, for any family $(p_t, e_t)_{0 \leq t < \tau}$ converging to $(p,e)$ as $t \uparrow \tau$, we have
		\begin{align}\label{eq v condition terminal in appendix}
\phi_-(p,e) \le \liminf_{t \rightarrow \tau}v(t,p_t,e_t) \le \limsup_{t \rightarrow \tau}v(t,p_t,e_t) \le \phi_+(p,e).
\end{align}
}
	\item{The limit $\lim_{t \uparrow \tau} v(t, P_t, E_t)$ exists and satisfies
			\begin{align}
			\label{terminal value inequality}
			\mathbb{P} \left[ \phi_{-} \left( P_\tau, E_\tau \right) \leq \lim_{t \uparrow \tau} v(s, P_t, E_t) \leq \phi_{+} \left( P_\tau, E_\tau \right) \right] = 1,
			\end{align}
		where $\phi_{-}$ and $\phi_{+}$ are, respectively, the left continuous and right continuous version of $\phi$, as defined by \eqref{phi minus plus definition}.}
	\item{ \label{comparison principle in appendix}
		[Comparison principle]
	For two different terminal conditions $\phi, \phi^{\prime} \in \cK$, denote by $v^{\phi}$ and $v^{\phi^{\prime}}$, respectively, the corresponding value functions. If $\phi$ and $\phi^{\prime}$ are such that $\phi(p,e) \geq \phi^{\prime}(p,e)$ for every $(p,e) \in \R^d \times \R$, then $v^{\phi}(t,p,e) \geq v^{\phi^{\prime}}(t,p,e)$ for every $t \in [0, \tau) \times \R^d \times \R$.	
}
	\end{enumerate}
\end{proposition}

\begin{theorem}
	\label{single period main theorem from thesis}
	Consider \eqref{general single period fbsde with p} and set $t_0 = 0$ to simplify notation (the analogous result holds for any $t_0 \in [0, \tau)$). Given any initial condition $(p,e) \in \mathbb{R}^{d} \times \mathbb{R}$, there exists a unique progressively measurable 4-tuple of processes $(P_t, E_t, Y_t, Z_t)_{0 \leq t \leq \tau} \in \mathcal{S}^{2,d}_{\mathrm{c}}([0,\tau]) \times \mathcal{S}^{2,1}_{\mathrm{c}}([0,\tau]) \times \mathcal{S}^{2,1}_{\mathrm{c}}([0,\tau)) \times \mathcal{H}^{2,d}([0,\tau])$ satisfying the dynamics in \eqref{general single period fbsde with p} with $(P_{0}, E_{0}) = (p,e)$ and such that
	\begin{align}
	\label{relaxed terminal condition with p in appendix}
	\mathbb{P} \left[ \phi_{-}(P_\tau,E_\tau) \leq Y_{\tau {-}} \leq \phi_{+}(P_\tau, E_\tau) \right] = 1,
	\end{align}
	where the functions $\phi_{-}$ and $\phi_{+}$ are the left and right continuous versions, respectively, of $\phi$, as defined in \eqref{phi minus plus definition}. 
	
	
	Finally, if $\sigma$ is bounded by $L$ then there exists a constant $C$ depending only on $L$, $L_{\phi}$ and $\tau$, such that $|Z_t| \leq C$ for almost every $(t, \omega) \in [0,\tau] \times \Omega$. In general, when $\sigma$ only satisfies the conditions in Assumption \ref{delarue assumptions}, then, given an initial condition $(p,e) \in \R^{d} \times \R$ and any $a \geq 1$, $Z$ satisfies
	\begin{align}
	\label{Z bound in single period theorem}
	\mathbb{E} \left[ \sup_{t \in [0,\tau]} |Z_t | ^{a}  \right] \leq C^{\prime} (1 + |p|^{a}),
	\end{align}
	for almost every $(t, \omega) \in [0,\tau] \times \Omega$, where $C^{\prime} > 0$ is a constant depending on $L$, $L_{\phi}$, $\tau$ and $a$.
\end{theorem}

We also have the following result, which was used in the proof of Proposition \ref{pr existence pricing function infinite model}, and is an extended version of Corollary 2.11 from \cite{carmona2013singularconservation}. Its proof follows along the same lines as the proofs in Section 2 of that paper. The full proof can be found in \cite{chotai2019thesis}. 
\begin{proposition}
	\label{v phi limit proposition more general from appendix}
	Let $(\phi_n)_{n \geq 1}$ be a sequence of terminal conditions for \eqref{general single period fbsde with p} in the sense of Theorem \ref{single period main theorem from thesis}. That is $\phi_n \in \cK$ for every $n$. For each $n$, denote by $L_{\phi_n}$ a Lipschitz constant for $\phi_n$. Suppose that the sequence $(\phi_n)_{n \geq 1}$ converges pointwise to a function $\phi$. Also assume that 
	\begin{align}
	\label{lipschitz constants uniformly bounded}
	\sup_{n \geq 1} L_{\phi_n} < \infty.
	\end{align} 
	Then, $\phi \in \cK$ also. For each $n \geq 1$, let $v^{\phi_n}$ and $v^{\phi}$ be the value functions, $v$, from Proposition \ref{v existence proposition from thesis} for \eqref{general single period fbsde with p} with terminal condition $\phi_n$ and $\phi$, respectively. Then $v^{\phi_n} \rightarrow v^{\phi}$ as $n \rightarrow \infty$. Convergence is uniform on compact subsets of $[0,\tau) \times \mathbb{R}^d \times \mathbb{R}$.
\end{proposition}

\end{document}